\documentclass[preprint]{aastex631}

\usepackage{amsmath}

\shorttitle{The combined effects of vertical and horizontal shear instabilities}
\shortauthors{Garaud et al.}
\graphicspath{{./}{figures/}}
\begin{document}

\title{The combined effects of vertical and horizontal shear instabilities in stellar radiative zones}

\author{Pascale Garaud}
\affiliation{Department of Applied Mathematics, Baskin School of Engineering, 1156 High Street, CA 95064 }

\author{Saniya Khan}
\affiliation{Institute of Physics, \'Ecole Polytechnique F\'ed\'erale de Lausanne (EPFL), Observatoire de Sauverny, 1290 Versoix, Switzerland}

\author{Justin M. Brown}
\affiliation{Department of Oceanography, Naval Postgraduate School, 1 University Circle, Monterey, CA, 93943}

\begin{abstract}
Shear instabilities can be the source of significant amounts of turbulent mixing in stellar radiative zones. Past attempts at modeling their effects (either theoretically or using numerical simulations) have focused on idealized geometries where the shear is either purely vertical or purely horizontal. In stars, however, the shear can have arbitrary directions with respect to gravity. In this work, we use  direct numerical simulations to investigate the nonlinear saturation of shear instabilities in a stably stratified fluid, where the shear is sinusoidal in the horizontal direction, and either constant or sinusoidal in the vertical direction. We find that, in the parameter regime studied here (non-diffusive, fully turbulent flow), the mean vertical shear does not play any role in controlling the dynamics of the resulting turbulence unless its Richardson number is smaller than one (approximately). As most stellar radiative regions have a Richardson number much greater than one, our result implies that the vertical shear can essentially be ignored in the computation of the vertical mixing coefficient associated with shear instabilities for the purpose of stellar evolution calculations, even when it is much larger than the horizontal shear (as in the solar tachocline, for instance). 
\end{abstract}


\section{Introduction} \label{sec:intro}

Quantifying vertical mixing by shear instabilities in stellar radiative zones is a long-standing question that dates back to the 1970s and the works of \citet{Zahn1974} \citep[see also][]{Schatzman1969,SpiegelZahn1970}. 
Indeed, shear is ubiquitous in stars. For instance, the gradual spin-down of the stellar surface by magnetized winds naturally generates some level of radial shear on a global scale via angular momentum transport \citep{Spadaal2016}. Similarly, large-scale meridional flows advecting the star's mean angular momentum poleward or equatorward naturally drive the emergence of a global-scale latitudinal shear \citep{Zahn1992,Spiegel1992}. More localized shear layers can be maintained in the vicinity of a differentially-rotating convection zone, as evidenced by the helioseismic discovery of the solar tachocline, which has both strong radial and latitudinal shear \citep{JCDSchou88,Thompson-etal96}. Internal gravity waves can in some cases amplify pre-existing shear and create strong localized radial shear layers in a star \citep{KumarQuataert1997,Ringot1998,KTZ1999,CharbonnelTalon2005}. 

Under the right conditions, the kinetic energy of the shear can drive and maintain turbulence via shear instabilities \citep{Richardson1920}. It is therefore important to understand what these conditions are and to quantify the resulting vertical mixing. This problem turns out to be particularly complex, however, because shear instabilities can take many different forms depending on the orientation of the shear with respect to the density stratification, as well as the strength and even the shape of the shear and density profiles \citep[cf.][]{DrazinReid2004,Caulfield2021}. Shear instabilities can also be affected by diffusion (viscous, thermal, or compositional), rotation and magnetic fields \citep{Chandrasekhar}. For these reasons, one must take a very gradual approach in studying the problem, adding only one new physical effect at a time to fully grasp its impact on the properties of the developing turbulence. 
Furthermore, while other instabilities can often be studied quite successfully using linear stability theory (i.e. stability to infinitesimal perturbations), this is rarely the case with shear instabilities. As long as viscosity is small, they can almost always be destabilized by suitably chosen finite-amplitude perturbations even when the background flow is linearly stable \citep[see, e.g.][]{Grossmann2000,Garaudal15,Avilaal2023}, and thus, the outcome sensitively depends on the initial conditions applied. As a result, studying the conditions for instability and resulting turbulent mixing often requires a combination of heuristic arguments and heavy-duty computational tools. 

For all of these reasons, progress in modeling mixing by shear instabilities in stars has been relatively slow, focusing for now on understanding simple geometries (vertical shear only, or horizontal shear only), ignoring the effects of magnetic fields (except in linear stability analyses) and mostly including the effect of rotation only insofar as it is the source of the shear in stars with differential rotation, but it is ignored thereafter. Within that limiting set of assumptions, however, starting with the work of \citet{Zahn1974} and \citet{Zahn1992}, combined with the more recent theoretical work and numerical experiments performed over the past 10 years (see the reviews of \citet{Lignieres2020} and \citet{Garaud2021}, as well as Sections \ref{sec:introvert} and \ref{sec:introhoriz} below), we now have a growing understanding of the criterion for instability. We also have models for the turbulent mixing associated with (non-rotating, non-magnetic) vertical and horizontal shear instabilities when these processes are considered independently. 

The main goal of this paper is to expand on this previous work to determine what happens to mixing by shear instabilities when vertical and horizontal shear are both present at the same time in a (non-rotating, non-magnetic) stably stratified region. With this in mind, we begin by briefly reviewing what is known about mixing by horizontal and vertical shear instabilities separately, which will also serve as a pedagogical introduction to the concepts needed for the interpretation of the red numerical simulations presented in this paper. 

\subsection{What is known about (non-rotating, non-magnetic) vertical shear instabilities}
\label{sec:introvert}

Loosely defined, vertical shear instabilities are processes that extract energy from the local vertical shear to generate and maintain some amount of vertical mixing. To do so in a stellar radiative zone, the shear must be strong enough to overcome the stabilizing effects of both viscosity and stratification. Because the viscosity is very small in stars, one could naively assume that it is never relevant -- and that is often, but not always the case (see below). Stratification, on the other hand, plays a dominant role in quenching vertical shear instabilities. Indeed, in order to tap into the mean shear kinetic energy reservoir, the turbulent eddies must move vertically but this incurs a corresponding energy transfer to the potential energy reservoir of the stratification when doing so adiabatically. 
The competition between the kinetic energy gain and potential energy cost is loosely captured by the so-called Richardson number $J$, defined here as the square of the ratio of the mean buoyancy frequency $N$ to the mean vertical shearing rate $S_m$, i.e.
\begin{equation}
 J = \frac{N^2}{S_m^2}.
     \label{eq:Rcrit}
\end{equation}
Assuming viscous and thermal diffusion are both negligible, the rate at which eddies can gain kinetic energy from the mean flow is proportional to $S_m^2$, while the rate at which they must provide potential energy to the stratification (by irreversibly lifting dense fluid parcels and lowering buoyant fluid parcels) is proportional to $N^2$.
 Vertical shear instabilities can therefore develop (linearly or nonlinearly) provided $J$ is (roughly) smaller than one \citep{Richardson1920}. A similar criterion that describes a necessary condition for the onset of linear instability can be derived more formally 
 \citep[cf.][]{Howard61}; in that case, the existence of an extremum in the shearing rate, or in other words, an inflection point in the flow profile, is generally also required \citep{Fjortoft1950,DrazinReid2004}. 
 
It is quite unusual, however, for stars to have extended regions that satisfy $J\le 1$ for long periods of time, except very close to the edge of a convective zone (where $N \rightarrow 0$), or when an external mechanism is present to sustain an intense shear \citep[such as accretion of material from a disk, see][]{MacDonald1983,Avilaal2023}. As such, the `dynamical' shear instabilities listed above do not appear to play a significant role in stellar evolution. 

All of the aforementioned arguments have assumed adiabatic perturbations, which is not necessarily the case if the stratification is primarily due to temperature and thermal diffusion is fast compared with the local dynamics. In the early 1970s, progress in understanding shear instabilities in the Earth's atmosphere \citep{Townsend1958}, where perturbations are not adiabatic due to radiative losses, led \citet{SpiegelZahn1970} to the realization that rapid thermal diffusion would have a similar effect in stellar interiors. \citet{Zahn1974,Zahn1992} then proposed the first model of thermally diffusive vertical shear instabilities (now often referred to as `secular' shear instabilities), where he argued that they mix momentum and chemical species vertically with a diffusivity 
\begin{equation}
D_{mix} = O\left(\frac{\kappa_T}{J} \right) 
\mbox{ as long as }
    JPr \ll C \mbox{ where } C \sim O(0.001) ,
    \label{eq:Zcrit}
\end{equation} 
and where $Pr = \nu/\kappa_T$ is the Prandtl number, $\nu$ is the viscosity and $\kappa_T$ is the thermal diffusivity. 
A series of direct numerical simulations (DNS hereafter) by \citet{Prat2013,Prat2014,Pratal2016}, as well as \citet{GaraudKulen16} and \citet{Garaud2017} confirmed the validity of (\ref{eq:Zcrit}) as long as $J \gg 1$ and thermal diffusion is important on the scale of the shear layer. 

In practice, however, $D_{mix}$ is always fairly small in secular shear instabilities \citep[at most, a couple of orders of magnitude larger than $\nu$, see][]{Garaud2021}. For this reason, they are not usually a significant source of mixing unless no other instabilities are present.

\subsection{What is known about (non-rotating, non-magnetic) horizontal shear instabilities}
\label{sec:introhoriz}

While vertical shear instabilities are either rare (in the dynamical regime) or not very efficient at mixing (in the secular regime), \citet{Zahn1992}  realized that horizontal shear instabilities could be a much more reliable source of mixing. Horizontal shear instabilities extract energy from the horizontal shear without any potential energy cost as long as the motions remain horizontal. As such, as long as magnetic fields and rotation are ignored, a horizontal shear is almost always unstable to purely horizontal perturbations \citep{BalmforthYoung2002,AroboneSarkar2012,Cope2020,Parkal2020}, that drive a meandering of the mean flow, and/or the formation of large-scale `pancake'-like vortices. The low viscosity then implies that these meanders and vortices decouple in the vertical direction, thus leading to the emergence of strong vertical shear on small vertical scales, even in the absence of any {\it mean} vertical shear. Small-scale vertical shear instabilities, and associated vertical mixing, can thus be triggered by the presence of an unstable large-scale horizontal shear. This general picture has been validated in DNS of \citet{Cope2020} and \citet{Garaud2020} (see more on this below). 

By contrast with the case of vertical shear instabilities where models generally agree with each other, and with the data from numerical simulations,
quantifying mixing by horizontal shear instabilities is a topic of ongoing investigations where no consensus has yet been reached. Several theoretical models have been put forward, starting with \citet{Zahn1992}, followed by the more recent works of \citet{Cope2020}, \citet{Garaud2020}, \citet{Lignieres2020},
\citet{Chinial2022}, \citet{Skoutnev2023} and \citet{Shah2023}. These models vary significantly in their detailed predictions because of differences in their respective founding assumptions and in the mathematical approach selected, and because their domains of validity are sometimes limited. It is not the purpose of this section to discuss each model in turn, as it would be difficult to do so without any bias \citep[see][for a rapid review of the subject]{Shah2023}. Instead, we summarize here what general agreement the models have, with the hope that these are concepts that can be relied upon with reasonable confidence. 

Generally speaking, all models attempt to relate the vertical mixing coefficient $D_{mix}$ to properties of the horizontal turbulence. \citet{Zahn1992}, for instance, directly related $D_{mix}$ to the horizontal turbulent diffusivity $\nu_h$, while the other models relate $D_{mix}$ to the assumed properties of large-scale horizontal eddies (which have characteristic large-scale horizontal velocity $U_h$ and lengthscale $L_h$). It is also generally agreed that as long as the flow is fully turbulent, $D_{mix} \propto l_z w_{rms}$ (where $l_z$ is a characteristic vertical eddy lengthscale and $w_{rms}$ is a characteristic vertical eddy velocity), and that both of these quantities must decrease with the increasing stratification. Finally, we know that there ought to be different regimes depending on whether the turbulent P\'eclet number based on $w_{rms}$ and $l_z$, namely $Pe_t =  w_{rms} l_z/ \kappa_T$, is larger than 1 (adiabatic regime) or smaller than 1 (diffusive regime), and that the transition between the two regimes is a function of the stratification and of the thermal diffusivity. It is also worth noting that viscosity is expected to become relevant once the stratification exceeds some threshold, although not all models take this into account. Because of this, the partitioning of parameter space into various regimes is relatively complex \citep[see, for instance,][]{Shah2023}. 

The main disagreements between the models emerge in the derivations of $w_{rms}$ and $l_z$ as functions of the properties of the horizontal turbulence. This also affects where the boundaries of the various regimes (adiabatic, diffusive, viscous) lie in parameter space. While the theoretical controversy is ongoing, there is some hope of gaining insight from DNS data. \citet{Cope2020} and \citet{Garaud2020} recently ran a series of DNS of horizontally sheared, stratified Kolmogorov flows at low Prandtl number. In these simulations, a constant-in-time body force ${\bf F} = F_0 \sin(k_h y) {\bf e}_x$ is applied to drive a vertically invariant, sinusoidal flow. The anticipated horizontal eddy scale is therefore an input parameter, $L_h =O(k_h^{-1})$, while the anticipated horizontal velocity scale is also an input parameter, obtained from balancing the nonlinear terms and the forcing in the streamwise component of the momentum equation: $U_h = O(\sqrt{F_0/ \rho_m k_h})$ where $\rho_m$ is the mean density of the fluid. \citet{Cope2020} explored cases where thermal diffusion is important at all scales, so the P\'eclet number based on the large scales, $Pe = U_h L_h / \kappa_T$, is smaller than 1. Meanwhile  \citet{Garaud2020} explored cases where thermal diffusion is negligible on the large scales ($Pe \gg 1$). 

In the thermally diffusive regime ($Pe < 1$), \citet{Cope2020} found, rather robustly, that there exists an intermediate regime of stratified turbulence where
\begin{equation}
D_{mix} \simeq w_{rms} l_z \propto \left( \frac{U_h  \kappa_T}{L_h^3 N^2} \right)^{1/2} U_h L_h . 
\label{eq:ZhDmix}
\end{equation}
This regime is valid as long as the  vertical scale of the turbulent eddies is much smaller than the domain size, but yet large enough to not be influenced by viscosity. The scaling law (\ref{eq:ZhDmix}) is consistent\footnote{assuming some flexibility in the interpretation of what $U_h$ and $L_h$ are} with the model predictions from \citet{Zahn1992}, \citet{Lignieres2020},  \citet{Skoutnev2023} and \citet{Shah2023} in their respective thermally-diffusive regimes, which is rather surprising as the scalings for $w_{rms}$ and $l_z$ individually differ in each model.  

However, the constraint $Pe < 1$ is rarely satisfied in stars. In the case where $Pe \gg 1$, \citet{Garaud2020} found that (\ref{eq:ZhDmix}) does not always apply and that two regimes exist depending on whether the turbulent P\'eclet number $Pe_t$ defined above is larger or smaller than one.  More specifically, $D_{mix}$ is independent of the thermal diffusivity when $Pe_t \gg 1$, but recovers the diffusive results of \citet{Cope2020} when $Pe_t \ll 1$. Unfortunately, as $Pe_t$ is an emergent quantity (rather than an input parameter), a model is needed to predict when the transition from non-diffusive to thermally diffusive turbulence occurs as a function of the input parameters. Here, the disagreement between the models themselves is significant; for instance, \citet{Skoutnev2023} finds that the solar tachocline would lie close to the regime transition while \citet{Shah2023} argue instead that it lies squarely in the non-diffusive regime. Finally, we note that running DNS in the regime $Pe \gg 1$ and $Pr \ll 1$ is numerically very challenging (as these require extremely large Reynolds numbers). As a result, numerical experiments have yet to reach sufficiently high values of $Pe$ to be able to test model predictions in that regime, and establish which one (if any) is correct.

\subsection{Outline of the proposed work}

While we will need to wait for future DNS to validate and/or invalidate existing models, we now return to the original question posed earlier, namely what happens when mean vertical and horizontal shear are combined, which is the more relevant scenario in stellar interiors. Indeed, the interplay between rotation, meridional circulations, spin-down by stellar winds, and shear instabilities naturally contributes to the maintenance of a large scale rotational shear in both radius and latitude, clarified in the seminal paper by \citet{Zahn1992}. The solar tachocline, for example, supports a large-scale horizontal differential rotation (with the equator rotating approximately 30 percent faster than the poles), as well as the well-known sharp vertical shear that is at the origin of its name \citep{Spiegel1992}. Red Giant Branch (RGB) stars are known to have substantial radial shear between the core and envelope \citep[see][and references therein]{Aerts2019} which must necessarily be associated with horizontal shear by Zahn's argument. Conversedly, stars with outer convective zones can exhibit substantial surface latitudinal differential rotation \citep{Reinholdal2013} which, by analogy with the solar tachocline, suggests the likely existence of strong radial shear below.

In this study, we continue past investigations by \citet{GaraudKulen16,Garaud2017}, \citet{Cope2020} and \citet{Garaud2020}, and run a series of DNS experiments which combine vertical and horizontal shear to address the effects of the former on the latter. Our goal is to establish a set of guiding principles to determine if and when one type of instability clearly dominates over the other, or when both interact in such a way that they cannot be modeled separately. We focus on the high P\'eclet number regime ($Pe \gg 1$) defined earlier, which is arguably the more relevant situation for the majority of stars \citep{Garaud2020}. 
Section \ref{sec:model} describes the model setup, in which two different vertical shear profiles (constant or sinusoidal) are tested. Section \ref{sec:analysis} presents our results, both qualitatively and then more quantitatively. Finally, Section \ref{sec:conclusions} summarizes our results while discussing important caveats of our model assumptions. 

\section{Model setup} \label{sec:model}

In what follows, we consider a stably-stratified (i.e., radiative) region of a star.
Gravity defines the vertical direction (${\bf g} = -g {\bf e}_z$), and the vertical extent of the domain is assumed to be much smaller than a density, pressure or temperature scaleheight. With this assumption, the local Brunt-V\"ais\"al\"a frequency $N$ can be assumed to be constant. 
We use the Boussinesq approximation for gases \citep{SpiegelVeronis1960}, noting that fluid flows in the deep interiors of stars are much slower than the local sound speed. A mean horizontal flow is driven by application of a body force ${\bf F} = F {\bf e}_x$, and/or the use of a shearing box (see below for more detail). We then study the development of shear instabilities on this mean flow, ignoring in this work the effects of rotation and magnetic fields. This assumption is made for simplicity and is briefly discussed in Section \ref{sec:conclusions}. Future work will investigate their effects separately.

The governing equations describing this model are
\begin{eqnarray}
\frac{\partial {\bf u}}{\partial t} + {\bf u}\cdot\nabla {\bf u} = - \frac{1}{\rho_m} \nabla p + \alpha g T {\bf e}_z + \nu \nabla^2 {\bf u} +  \frac{F}{\rho_m}  {\bf e}_x , \label{eq:momentumdim}\\ 
\nabla \cdot {\bf u = 0},  \label{eq:boussinesqdim} \\
\frac{\partial T}{\partial t} + {\bf u}\cdot\nabla T + \frac{N^2}{\alpha g}w = \kappa_T \nabla^2 T ,
\label{eq:temperaturedim}
\end{eqnarray}
where ${\bf u} = (u,v,w)$ is the velocity field, $T$ is the potential temperature fluctuation away from the stably-stratified background state, and $p$ is the pressure fluctuation away from hydrostatic equilibrium. In addition, $\rho_m$ is the mean density of the fluid, $\alpha$ is the coefficient of thermal expansion, $\nu$ is the viscosity, and $\kappa_T$ is the thermal diffusivity. All of these fluid properties are assumed to be constant in the small region considered. The body force ${\bf F} = F {\bf e}_x$ varies sinusoidally in the $y$ direction, with a horizontal wavenumber $k_h = 2\pi/L_y$ where $L_y$ is the domain width. 

We then non-dimensionalize the equations as in \citet{Cope2020} and \citet{Garaud2020}. We use $k_h$ as the inverse unit length. Anticipating a balance between the forcing and the inertial terms in the horizontal component of the momentum equation, we define
\begin{equation}
U_F = \left( \frac{F_0}{\rho_m k_h} \right)^{1/2} \rm \, ,
\label{eq:uf}
\end{equation}
where $F_0$ is the amplitude of the sinusoidal body force, and use $U_F$ as the unit velocity. The unit time is then defined as $(k_h U_F)^{-1}$ and the unit temperature as $N^2 / \alpha g k_h$. Note that since the lengthscale used in (\ref{eq:uf}) is $k_h^{-1}$, this non-dimensionalization more appropriately captures the scalings of anticipated horizontal shear instabilities than of vertical shear instabilities (which would likely care about the vertical lengthscale of the shear or the domain height $L_z$).

The non-dimensional governing equations corresponding to (\ref{eq:momentumdim})-(\ref{eq:temperaturedim}) are then:
\begin{eqnarray}
\frac{\partial \hat{\bf u}}{\partial t} + \hat{\bf u}\cdot\nabla \hat{\bf u} = - \nabla \hat{p} + B \hat{T} {\bf e}_z + Re^{-1} \nabla^2 \hat {\bf u} +   \hat F {\bf e}_x, \label{eq:momentum} \\
\nabla \cdot \hat{\bf u} = 0, \label{eq:boussinesq} \\
\frac{\partial \hat{T}}{\partial t} + \hat {\bf u}\cdot\nabla \hat{T} + \hat{w}  = Pe^{-1} \nabla^2 \hat{T}. \label{eq:temperature} 
\end{eqnarray}

In that follows, hatted quantities are non-dimensional, while those without hats are dimensional, except for the independent variables $x,y,z,t$ and associated operators which are assumed to be non-dimensional (hats in this case are removed to avoid crowding the notation). Several parameters are introduced in these equations: the domain-scale Reynolds number $Re$, which is the ratio of the viscous diffusion timescale to the turbulent advection timescale; the domain-scale Péclet number $Pe$, which is the ratio of the thermal diffusion timescale to the turbulent advection timescale; and the buoyancy parameter $B$, which is the square of the ratio of the Brunt-Väisälä frequency $N$ to the horizontal shearing rate $k_h U_F$ of the anticipated mean flow:
\begin{equation}
Re = \frac{U_F}{k_h \nu} \, , \quad Pe = \frac{U_F}{k_h \kappa_T}  \quad \mbox{and } B = \frac{N^2}{k_h^2 U_F^2} \rm \, .
\end{equation}

As we are interested in studying the {\it combination} of vertical and horizontal shear on the development of shear instabilities, we note that there are different possible ways of driving the shear. In this work, we always drive the horizontal shear using a body force as in \citet{Cope2020} and \citet{Garaud2020}, so the results can be compared with their work in the limit where the vertical shear is zero. To apply vertical shear, however, there are several options. One of them is to use the standard shearing box formalism \citep[as in, e.g.,][]{Pratal2016} and drive a constant vertical shear. Another option is to use a body force that is sinusoidal in both horizontal and vertical directions. Both types of shear may be expected in stars. For instance, the constant shear case would be appropriate to model stars where the vertical shear varies slowly with depth. On the other hand, the sinusoidal shear case could be appropriate to model stars with complex vertical shear profiles, such as the ones that can be driven by the nonlinear interaction between gravity waves and differential rotation.

In what follows, we use and compare both options, which serves two purposes. First, it might help establish the role of the vertical shear lengthscale on the emergent turbulence. In the sinusoidal model proposed below, the vertical shear lengthscale is finite while it is infinite in the constant shear case. Second, and likely more importantly,
a sinusoidal vertical shear has an inflection point, and can excite a linear vertical shear instability provided the Richardson number is small enough. In the constant shear case, by contrast, the vertical shear is not (on its own) linearly unstable even at low Richardson number. One might therefore expect to obtain rather different outcomes with and without the inflection point. 

These two different options require two different codes, but are both based on the PADDI code that was already used in the works of \citet{Cope2020} and \citet{Garaud2020}. PADDI is a high-performance, triply-periodic pseudo-spectral DNS code, described in \citet{Traxleretal2011b} (see their Appendix A). 

\subsection{Sinusoidal shear in the horizontal direction and uniform shear in the vertical direction}
\label{sec:linear}

In a first model setup, we apply the same non-dimensional body force in equation (\ref{eq:momentum}) as in \citet{Cope2020} and \citet{Garaud2020}, namely, 
\begin{equation}
{\bf \hat{F}} = \hat F {\bf e}_x = \sin(y) {\bf e}_x.
\end{equation}
In addition, we drive a constant vertical shear using the shearing-box formalism. 
The numerical methodology is, overall, quite similar to the PADDI code, and is described in detail in \citet{Brown2021}. In this formalism, the simulation grid deforms over time, following the flow of the (prescribed) uniform non-dimensional vertical shear, $\hat S$, which is described by the following coordinate transformation:
\begin{equation}
    x' =x+\hat Szt, \quad 
    y'=y,  \quad
    z'=z, \quad
    t'=t.
\end{equation}
The Fourier modes of the simulation are expanded as complex exponentials in the primed coordinate system, and thus, in addition to horizontal periodicity in $x$ and $y$, the fields are assumed to be periodic in $z'$ instead of $z$.
When the inclination of the grid, $\partial z'/\partial x'$ exceeds $\hat L_z/2\hat L_x$, the simulation is remapped to a new coordinate system with an inclination of $\partial z''/\partial x''=\partial x'/\partial z' - L_z/L_x$.
This transformation is performed with the appropriate dealiasing \citep{Delorme1985}.
Transforming the system in this way ensures minimal cell deformation and retains periodicity in the new coordinate system. Note that in these simulations, the velocity field $\hat {\bf u}$ represents the flow perturbations around a state of constant shear, rather than the full flow field. % This is correct

In all simulations, the non-dimensional size of the domain is fixed to $\hat L_x=4\pi$ and $\hat L_y=\hat L_z=2\pi$. Selecting a domain that is at least twice a long in the $x$ direction as in the $y$ direction ensures that it contains at least one wavelength of the fastest-growing mode of the primary horizontal shear instability \citep{Cope2020}. The resolution is fixed to have 384 equivalent grid points per interval of length $2\pi$ in each direction, which ensures that they are all well-resolved at the parameters selected (see below).

The initial conditions for the simulation are either $\hat u(x,y,z,0) = \sin(y) + \hat \mu(x,y,z)$ where $\hat \mu$ is some small amplitude white noise, or, the simulation is continued from the end of another simulation at different values of $B$ (this greatly helps reduce compute time). The final statistically stationary state achieved by the simulations seems to be independent of the initial conditions, at least for the cases run here\footnote{It is quite likely that different final states could exist if the initial temperature profile is not linear.}. 

\subsection{Sinusoidal shear in the horizontal and vertical directions}
\label{sec:sinusoidal}

In a second model setup, we force the mean shear with a body force that varies sinusoidally in both horizontal and vertical directions, such that 
\begin{equation}
{\bf \hat{F}} = \hat F {\bf e}_x = \sin(y+\hat Sz){\bf e}_x.
\end{equation} 

\begin{figure}[!h]
  \centering
  \resizebox{1.01\hsize}{!}{\includegraphics[angle=-0]{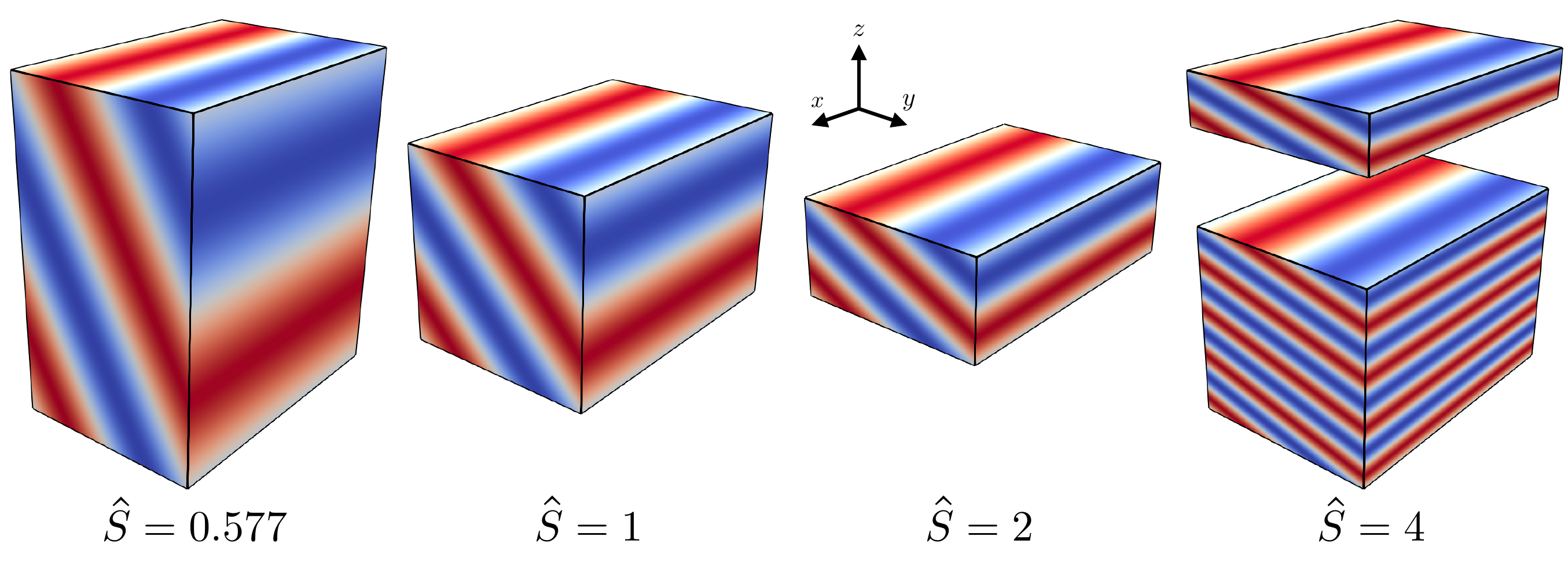}}
  \caption{Illustration of the sinusoidally-sheared model setup showing the forcing profile $\propto \sin(y+\hat S z)$, in the different configurations: $\hat S = 0.577, 1, 2, 4$. In each case, $\hat L_z = 2\pi/\hat S$ to ensure periodicity. For the $\hat S= 4$ case, additional simulations were run with $\hat L_z = 2\pi$, hence both are shown here.}
\label{fig:geometry}
\end{figure}

As in the case with constant vertical shear described above, the non-dimensional horizontal dimensions of the domain are $\hat L_x = 4\pi$ and $\hat L_y = 2\pi$. To ensure periodicity in the vertical direction, we choose $\hat L_z = 2\pi/ \hat S$ unless otherwise specified. The resolution is 384 equivalent grid points per interval of length $2\pi$ in each direction, as above, so the results of the constant vertical shear case and the sinusoidal vertical shear case can easily be compared to one another. For the largest shearing rate considered in our simulations, namely $\hat S = 4$, taking $\hat L_z = \pi/ 2$ results in a rather short vertical domain. Worrying that this could artificially impact the results, we also ran simulations with $\hat L_z = 2\pi$, but found that the time-averaged mean properties of the flow extracted in the shorter and taller domain sizes are very similar (see Table \ref{tab:results}). We therefore kept the larger $\hat L_z = 2\pi$ runs for the $\hat S = 4$ case (since they were available) but were satisfied with a shorter domain height $\hat L_z = \pi$ for the $\hat S = 2$ simulations. 

The initial conditions for the simulations are either $\hat u(x,y,z,0) = \sin(y+\hat Sz) + \hat \mu(x,y,z)$ or the simulation is restarted from the end of another simulation at different values of $B$. Again, the final statistically stationary state achieved appears to be independent of the initial conditions with the same caveat as above.

\section{Results} \label{sec:analysis}

\subsection{Simulations parameters}
\label{sec:params}

In what follows, we present a large suite of DNS of horizontally and vertically sheared stratified flows. In almost all cases, we have run simulations at the same parameters for the case of constant and sinusoidal vertical shear, which enables us to distinguish between their effects in controlling the properties of the turbulence. The results presented have a Reynolds number $Re = 600$, which is the largest value we can realistically achieve while at the same time running a large number of simulations. As shown by \citet{Garaud2020} in the case without vertical shear, $Re = 600$ is large enough to ensure that the effects of viscosity on the flow dynamics are essentially negligible, except for the largest stratification selected below ($B = 100$), where they are partially suppressing the turbulence, leading to spatial intermittency in the turbulence intensity. The Prandtl number is fixed to be $Pr = 0.1$, which is substantially smaller than one, while at the same time ensuring that the P\'eclet number based on the large-scale horizontal flow remains very large ($Pe = 60$). This parameter regime, where both $Re$ and $Pe$ are large, but $Pr$ is small, is consistent with what one would expect in stellar interiors, but the numerical values achieved are much less extreme than in reality \citep[where $Re$ and $Pe$ are generally many orders of magnitude larger, and $Pr$ is many orders of magniture smaller, see][]{Garaud2020}.

Having fixed $Re = 600$, $Pe = 60$, we then ran a grid of simulations for $B \in \{1, 10,30,100\}$ and $\hat S \in \{0,0.577,1,2,4\}$. The range of $B$ was selected to span regimes where the turbulence is effectively unstratified ($B =1$), where stratification controls but does not inhibit the turbulence ($B = 10$ and $B=30$) and where stratification is sufficiently strong to  intermittently suppress the turbulence ($B=100$); see \citet{Cope2020} and \citet{Garaud2020} for a description and characterization of these various regimes. Note that the threshold values of $B$ that respectively delimit the three regimes (when $\hat S =0$) depend on the Reynolds and P\'eclet number, so the choices made here are only valid for $Re = 600$, $Pe=60$ \citep[see][for a more complete map of parameter space]{Shah2023}. The range of $\hat S$ was selected to span regimes of zero and weak vertical shear ($\hat S = 0.577$), moderate vertical shear ($\hat S = 1$, for which the typical horizontal and vertical shear are the same) and large vertical shear (up to $\hat S = 4$). We did not run simulations for $\hat S$ larger than $4$, as these require much smaller timesteps and become too computationally expensive. The model parameters, and salient results, are presented in Table 1. 

This choice of parameters implies that the vertical shear instabilities expected of the mean shear alone would be of the dynamical kind rather than the diffusive kind, should they take place. Indeed, we can define a P\'eclet number based on the vertical shear to be 
\begin{equation}
Pe_S = \frac{SL_S^2}{\kappa_T}, 
\end{equation}
where $S$ is the dimensional shearing rate, and $L_S$ is the dimensional vertical shear lengthscale. In the models described above, $L_S = (\hat S k_h)^{-1}$ for the sinusoidally-sheared case (so $Pe_S = Pe / \hat S$), and $L_S$ is technically infinite in the constant shear case, but in practice is related to size of the domain $L_z$ (so $Pe_S = \hat S Pe$). With the choice of parameters selected here, $Pe_S$ is always much greater than one. Based on the work of \citet{GaraudKulen16}, we then anticipate that instabilities of the mean applied vertical shear would drive adiabatic motions rather than thermally diffusive ones in the absence of horizontal shear. 

We now define the quantity $J_{in}$ as the `input' Richardson number based on the imposed vertical shear:
\begin{equation}
J_{in} = \frac{N^2}{S^2} = \frac{B}{\hat S^2}.
\end{equation}
This quantity is listed in Table 1 for each simulation. As explained in Section \ref{sec:introvert}, when the mean Richardson number is significantly smaller than 1, adiabatic perturbations gain more energy from the mean vertical shear than they lose from mixing the stratification \citep{Richardson1920}. Turbulence can be sustained in this manner, and we therefore expect the vertical shear to play an important role in the flow dynamics. By contrast, when the Richardson number (and the P\'eclet number) are both significantly greater than 1, turbulence cannot be sustained by the mean vertical shear only, and one must rely on the horizontal shear instabilities instead to drive it. 

Note that most stars generally have $J \gg 1$ in their stably stratified radiative zones, even in regions of strong vertical shear (in the bulk of the solar tachocline, for instance, $J \sim 10^3 - 10^4$ and only drops to zero very close to the base of the convection zone). Nevertheless, for the sake of completeness, we have run simulations that span a wide range of values of $J_{in}$ including cases where $J_{in} < 1$. 

\begin{table}[]
        \label{tab:results}
\centering
\vspace{0.3cm}
\begin{tabular}{ccccccccc}
\tableline
      $B$  & $\hat S$ & $J_{\rm in}$ & $\hat u_{rms}$ &  $\hat w_{rms}$ & $\hat l_z$     & $\hat u_{rms}$ &   $\hat w_{rms}$ & $\hat l_z$                    \\
\tableline
$1$ & $0^{(a)}$ & $\infty$ & $1.69 \pm 0.3$ &  $0.87 \pm 0.07 $ & $0.48 \pm 0.06$  &  &    \\
$1$ & $0.577$& $3$ & $1.77\pm0.31$ &  $0.94 \pm 0.07 $ & $0.51 \pm 0.07$ &   &\\
$1$ & $1$ & $1$ & $2.11\pm0.24$& $1.01 \pm 0.07$ & $0.44 \pm 0.05$ & $1.89 \pm 0.15$ & $0.95 \pm 0.06$  &  $0.46 \pm 0.06 $\\
$1$ &$2$& $0.25$  & $2.47\pm0.25$& $1.24 \pm 0.08$ & $0.40 \pm 0.06$  & $1.4 \pm 0.08$ & $0.79 \pm 0.04$  & $0.42 \pm 0.06$\\
$1$ & $4$& $0.0625$  & $4.42\pm0.4$& $2.88 \pm 0.31$ & $0.51 \pm 0.13$  &  $0.97 \pm 0.03$ & $0.58 \pm 0.02$ & $0.32 \pm 0.03$\\
$1$ & $4^{(b)}$& $0.0625$  & & &  & $0.99 \pm 0.01$  & $0.59 \pm 0.007$ & $0.30 \pm 0.01$ \\
\tableline
$10$ & $0^{(a)}$ & $\infty$& $2.17 \pm 0.2$ & $ 0.64 \pm 0.07  $ & $ 0.29\pm 0.02$ & & & \\
$10$ & $0.577$ & $30$  & $1.59\pm0.22$&  $0.60 \pm 0.06 $ & $0.30 \pm 0.03$  & $1.50\pm0.12$&$0.59\pm 0.03$  & $0.32 \pm 0.02$\\
$10$ & $1$ & $10$ & $1.95\pm 0.16$& $0.67 \pm 0.06$ & $0.30 \pm 0.02$   & $1.43\pm 0.09 $& $0.59\pm0.03$ & $0.31 \pm 0.01$ \\
$10$ & $2$& $2.5$  & $1.83\pm 0.24$& $0.62 \pm 0.06$ & $0.32 \pm 0.04$  &$2.72 \pm 0.15$ & $0.78 \pm 0.09$  & $0.22 \pm 0.04$\\
$10$ & $4$ & $0.625$ & $2.48\pm 0.11$& $0.75 \pm 0.05$ & $0.18 \pm 0.007$  & $1.64\pm 0.06$&$0.55 \pm 0.04$  & $0.20 \pm 0.01 $\\
$10$ & $4^{(b)}$& $0.625$  & &  &   & $1.64\pm 0.04$  & $0.55 \pm 0.02$ & $0.20 \pm 0.01$ \\
\tableline
$30$ & $0^{(a)}$& $\infty$ & $2.53 \pm 0.25$& $ 0.46 \pm 0.08  $ & $ 0.20\pm 0.01$  &&& \\
$30$  & $0.577$ & $90$ & $1.89\pm0.17$& $0.44 \pm 0.07 $ & $0.20 \pm 0.01$ & $ 1.66 \pm 0.13$  & $0.42 \pm 0.04$   & $0.21 \pm 0.01$\\
$30$ & $1$ & $30$ & $1.73\pm0.20$& $0.42 \pm 0.05$ & $0.21 \pm 0.01$   & $1.52 \pm 0.15$ &$0.41 \pm 0.04$   & $0.21 \pm 0.01$\\
$30$ & $2$ & $7.5$ & $1.81\pm0.25$& $0.49 \pm 0.08$ & $0.21 \pm 0.01$ & $1.23 \pm 0.13$& $0.39 \pm 0.02$   & $0.22 \pm 0.01$\\
$30$ & $4$& $1.875$  &$2.21\pm0.27$ & $0.46 \pm 0.04$ & $0.20 \pm 0.02$ & $2.52\pm 0.17$& $0.55 \pm 0.13$ & $0.18 \pm 0.05$\\
$30$ &  $4^{(b)}$ & $1.875$ &  &  &  & $2.65\pm 0.05$ &   $0.48 \pm 0.02$ & $0.13 \pm 0.008$ \\
\tableline
$100$ & $0^{(a)}$& $\infty$  & $1.94\pm 0.15$  &  $ 0.26 \pm 0.06  $ & $ 0.15\pm 0.01$ &&& \\
$100$ & $0.577$ & $300$ & $1.87\pm0.13$ &  $0.26 \pm 0.04 $ & $0.16 \pm 0.01$ & $1.82\pm 0.06$ & $0.25 \pm 0.02$  & $0.15 \pm 0.06$\\
$100$ & $1$ & $100$ &$1.58\pm0.17$& $0.25 \pm 0.03$ & $0.16 \pm 0.02$ & $1.95\pm 0.11$ & $0.26 \pm 0.04$ & $ 0.15 \pm 0.001$\\
$100$ & $2$  & $25$ & $2.05\pm0.15$& $0.29 \pm 0.04$ & $0.17 \pm 0.01$ & $1.78\pm0.15$& $0.25 \pm 0.05$ & $0.15 \pm 0.007$\\
$100$ & $4$  & $6.25$ & $1.61\pm0.13$& $0.25 \pm 0.04$ & $0.25 \pm 0.05$ & $1.41\pm0.28$&$0.21 \pm 0.05$ & $0.14 \pm 0.01$\\
$100$ & $4^{(b)}$ & $6.25$  &&  &    & $1.31\pm0.08$& $0.19 \pm 0.02$ & $0.14 \pm 0.04$ \\
\tableline
\end{tabular}
\caption{Input parameters and salient results for the constant vertical shear case (columns 4-6) and the sinusoidal shear case (columns 7-9). In all of the simulations, $Re = 600$, $Pe = 60$. The rms velocities $\hat u_{rms}$ and $\hat w_{rms}$, and the vertical eddy scale $\hat l_z$, are extracted from the DNS as described in Section \ref{sec:resquant}. $(a)$ denotes simulations from \citet{Garaud2020}. $(b)$ denotes sinusoidal shear simulations in a domain of height $\hat L_z = 2\pi$ rather than $\hat L_z = 2\pi/\hat S$. }
\end{table}

\subsection{Qualitative characteristics of the flow evolution and properties}
\label{sec:resqual}

\subsubsection{Temporal evolution of selected simulations}
\label{sec:qualtemp}

\begin{figure}[!h]
  \centering
  \resizebox{.9\hsize}{!}{\includegraphics[angle=0]{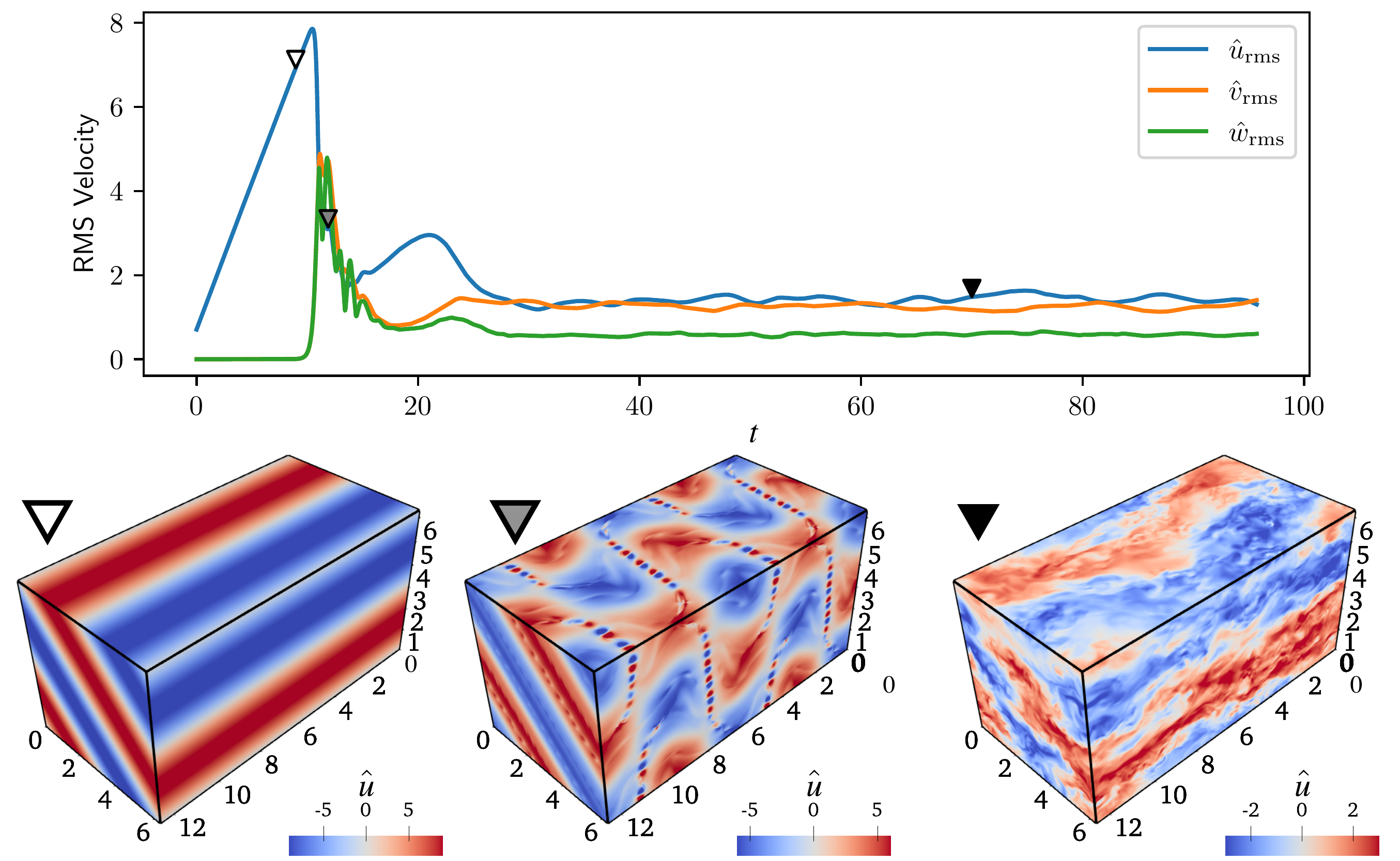}}
  \caption{Top: Rms velocity components $\hat u_{rms}$, $\hat v_{rms}$, and $\hat w_{rms}$ as a function of time, for a simulation with $Re=600, Pe=60, B=10, \hat S =1$ where the vertical shear is sinusoidal. Bottom:  Snapshots of the streamwise velocity $\hat u$ at selected times ($t = 9$, $t = 12$ and $t= 70$, see triangular symbols for corresponding times in top panel) showing the mean flow before onset of instability, the onset of instability, and the saturated state, respectively. }
\label{fig:B10S1}
\end{figure}

In this section we describe the typical temporal evolution and route to nonlinear saturation of the turbulence for selected simulations to illustrate the range of possible behaviors of the two model systems. We begin by looking in Figure \ref{fig:B10S1} at a simulation which has a relatively large input Richardson number, anticipating that it should be dominated by horizontal shear instabilities. It has a mean sinusoidal shear in both $y$ and $z$ directions, with $B=10,   \hat S = 1$ (so $J_{in} = 10$), and is initialized from noise at $t=0$. The top panel shows as solid lines the quantities $\hat u_{rms}(t)$, $\hat v_{rms}(t)$ and $\hat w_{rms}(t)$, defined as 
\begin{equation}
    \hat u_{rms}(t) = \langle \hat u^2 \rangle^{1/2}, \quad \hat v_{rms}(t) = \langle \hat v^2 \rangle^{1/2},  \mbox{ and } \hat w_{rms}(t) = \langle \hat w^2 \rangle^{1/2},
\end{equation}
where the brackets denote a volume average. Snapshots of the streamwise velocity field at selected times are shown in the bottom panels. 

At early times, the mean flow driven by the body force $\hat {\bf F} = \sin(y+\hat Sz){\bf e}_x$ simply grows linearly with time (see first snapshot). This flow is linearly unstable to shear instabilities, and perturbations begin to grow. They reach a sufficiently large amplitude to disrupt that mean flow around $t = 10$. The perturbations are spatially complex (see second snapshot), and their nature depends quite strongly on the dominant instability at any given point in time. While they are interesting in their own right, they are not the subject of this investigation as they rapidly evolve into fully-developed turbulence that saturates into a statistically stationary state (see third snapshot). 

For this particular choice of input parameters, the statistically stationary state achieved looks quite similar to the one described in \citet{Garaud2020} for simulations with no vertical shear. A horizontal shear instability of the mean flow causes and maintains large-scale horizontal meanders. After saturation, these are characterized by typical horizontal velocity components that are of the same order of magnitude, i.e., $\hat u_{rms} \simeq \hat v_{rms}$. The meanders decouple vertically on some short vertical lengthscale, and the resulting streamwise flow looks quite different from the one that is directly forced by $\hat {\bf F}$ (this can be seen by comparing the first and third snapshots). This decoupling generates substantial {\it emergent, small-scale} vertical shear that seems unstable to vertical shear instabilities even though the mean shear is not and therefore drives vertical mixing on small scales. The strong stratification implies that the typical vertical velocities are much smaller than the horizontal ones, so $\hat w_{rms} \ll \hat u_{rms},\hat v_{rms}$.

\begin{figure}[!h]
  \centering
  \resizebox{.9\hsize}{!}{\includegraphics[angle=0]{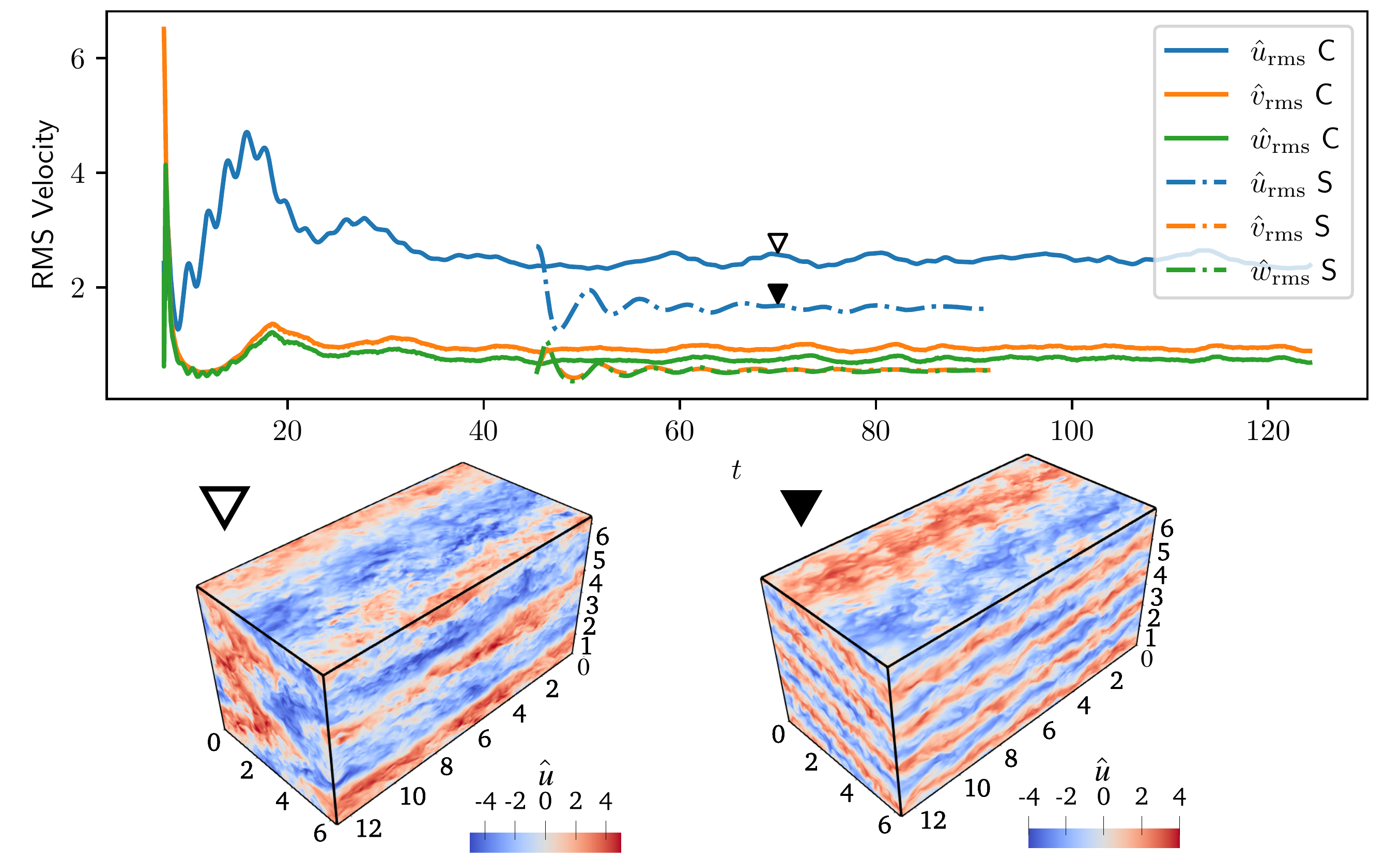}}
  \caption{Top: Rms velocity components $\hat u_{rms}$, $\hat v_{rms}$, and $\hat w_{rms}$ as a function of time, for simulations with $Re=600, Pe=60, B=10, \hat S =4$. The solid lines show the constant shear case, while the dashed lines show the sinusoidal shear case. Bottom left: Snapshot of $\hat u$ at $t= 70$ in the constant shear case. Bottom right: Snapshot of $\hat u$ at $t= 70$ in the sinusoidal shear case. }
\label{fig:B10S4}
\end{figure}

Figure \ref{fig:B10S4}, by contrast, shows two simulations that have a relatively low input Richardson number, namely $J_{in} = 0.625$ ($B = 10$, $\hat S = 4$). Both were restarted from other simulations at the same shearing rate but a different stratification. The solid lines show the case where the background shear is constant, while the dashed lines show the case where the background shear is sinusoidal. This time, we see that $\hat v_{rms}$ is substantially smaller than $\hat u_{rms}$ in both simulations once in the statistically stationary state. We find that $\hat w_{rms} \sim \hat v_{rms} \ll \hat u_{rms}$, consistent with the findings of \citet{GaraudKulen16} and \citet{Garaud2017} for cases without horizontal shear. The two simulations are otherwise quite different, however. In the sinusoidal shear case, we see in the corresponding snapshot that the streamwise velocity field $\hat u$ is dominated by a coherent mean flow $\propto \sin(y+\hat Sz)$. This implies that it contains the necessary inflection points to directly trigger a vertical shear instability since $J_{in}$ is small.  The constant shear case, as discussed earlier, is linearly stable to the vertical shear instability. However, the emergence of meanders from the horizontal shear instability rapidly creates the vertical structure  needed to trigger the vertical shear instability (see more on this in the next section). Because the meanders are necessary for saturation, the velocity components satisfy $\hat w_{rms} \le \hat v_{rms} \ll \hat u_{rms}$ in this case. 

\begin{figure}[!h]
  \centering
  \resizebox{.9\hsize}{!}{\includegraphics[angle=0]{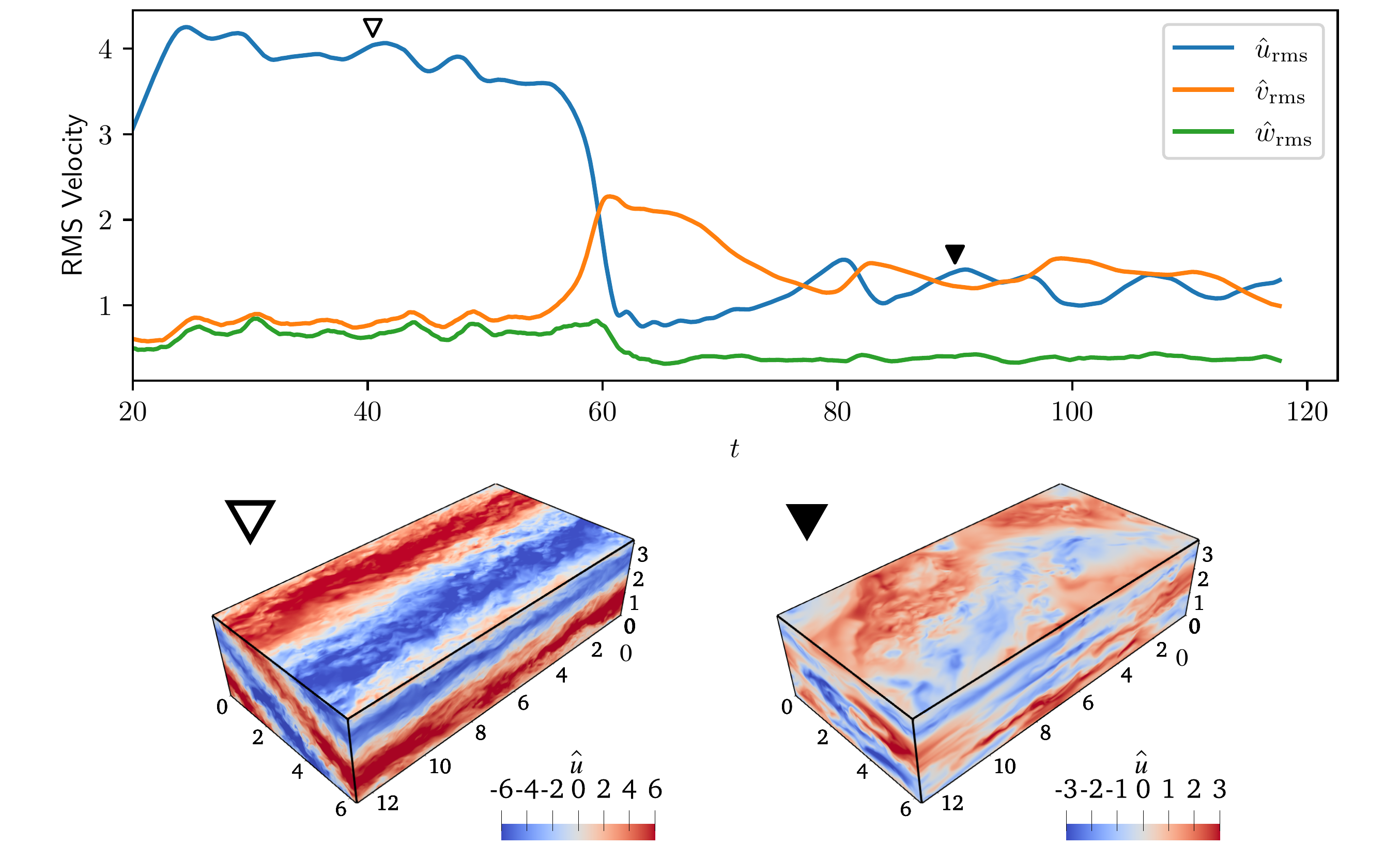}}
  \caption{Top Rms velocity components $\hat u_{rms}$, $\hat v_{rms}$, and $\hat w_{rms}$ as a function of time, for a simulation with $Re=600, Pe=60, B=30, \hat S =2$ where the vertical shear is sinusoidal. Botton: Snapshots of the streamwise velocity $\hat u$ at selected times ($t = 40$, and $t= 90$, see corresponding triangular symbols in the top panel.).}
\label{fig:B30S2}
\end{figure}

Finally, we note that while most simulations achieve a statistically stationary state relatively rapidly, in a few instances we have observed a long-lived transient flow dominated by vertical shear instabilities, which eventually transitions into a statistically stationary state that is dominated by horizontal shear instabilities. This can be seen in the case presented in Figure \ref{fig:B30S2}, which has an applied mean sinusoidal shear in both the $y$ and $z$ directions, with $B=30, \hat S = 2$, and was initialized at $t \simeq 20$ from a corresponding simulation with a lower stratification ($B=10$) and identical shearing rate. At early times, the turbulence has properties that are characteristic of simulations dominated by vertical shear instabilities (with $\hat w_{rms} \simeq \hat v_{rms} \ll \hat u_{rms}$). This state lasts for about 40 time units, and appears to be relatively stationary in time. However, around $t=60$ a complete reorganization of the flow takes place. We see that $\hat v_{rms}$ increases rapidly while $\hat u_{rms}$ decreases, owing to the growth and saturation of a large-scale horizontal shear instability. A new (and this time true) statistically stationary state is established, whose properties are qualitatively similar to the ones described earlier for the $B=10$, $\hat S=1$ case. Notably, we see that once again $\hat w_{rms} \ll \hat u_{rms} \simeq \hat v_{rms}$. 

Based on these results, we have run all simulations for at least 50 time units once it appears that a statistically steady state has been reached (and often for much longer). 

\subsubsection{Qualitative properties of the statistically stationary state}
\label{sec:qualsss}

\begin{figure}[!h]
  \centering
  \resizebox{0.85\hsize}{!}{\includegraphics[angle=-0]{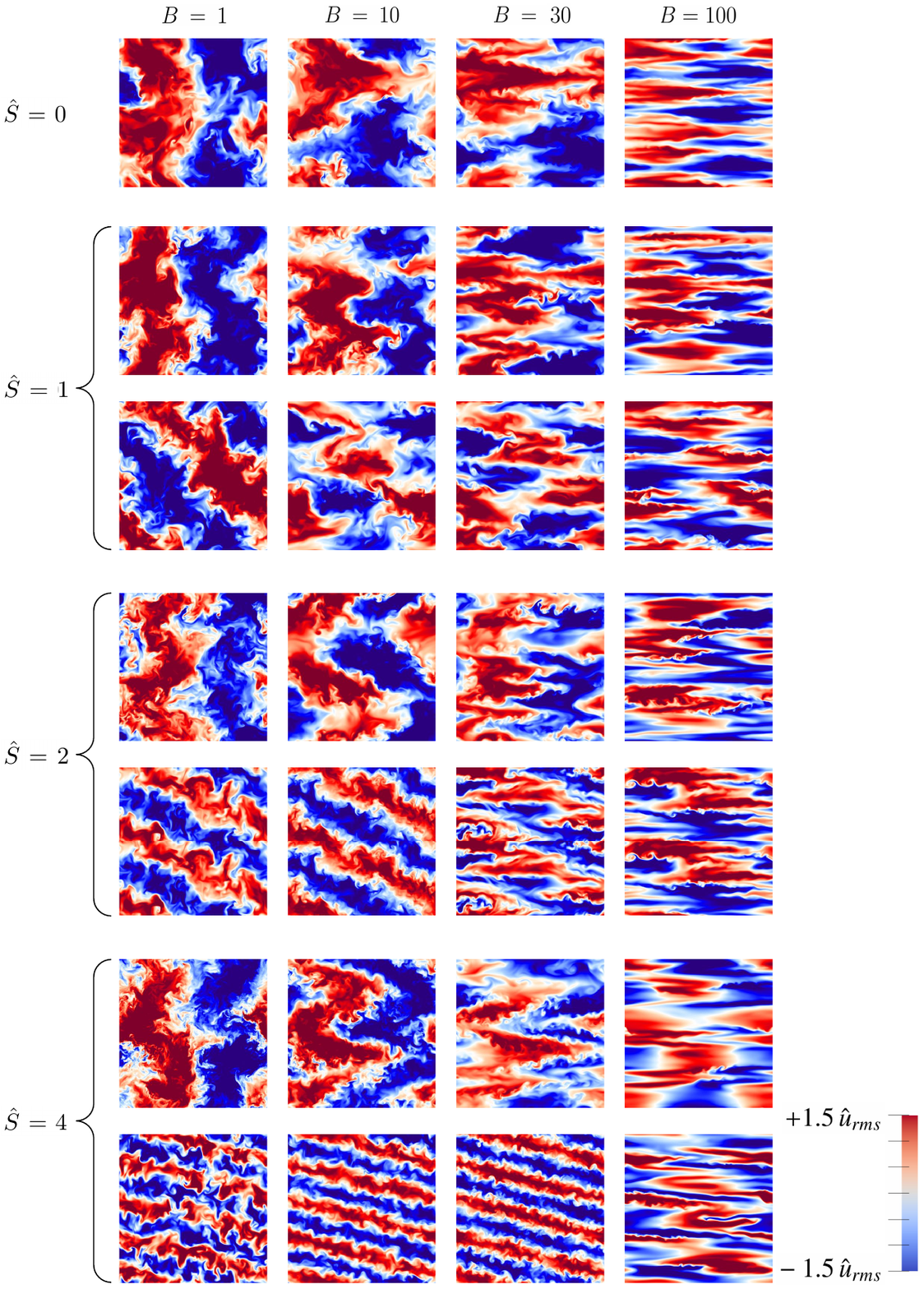}}
  \caption{Simulation snapshots of $\hat u$ in the $x=0$ plane, for increasing shear $\hat S$ (from top to bottom) and stratification $B$ (from left to right). Simulations with $\hat S = 0$ are  from \citet{Garaud2020}. For $\hat S \in \{1,2,4\}$, the top row shows snapshots for the constant vertical shear case (Sect. \ref{sec:linear}), and the bottom one for the sinusoidal shear case (Sect. \ref{sec:sinusoidal}). The color bar is in units of $\hat u_{rms}$ given for each simulation in Table \ref{tab:results}. All simulations were run in a domain with height $\hat L_z= 2\pi$ except for the sinusoidally-forced $\hat S = 2$ case, which was run in a domain with $\hat L_z= \pi$ (two copies of the snapshot are thus shown). }
\label{fig:snapshots}
\end{figure}

We now investigate visually the properties of the statistically stationary state ultimately achieved by each of the simulations. Figure \ref{fig:snapshots} shows typical snapshots of the streamwise component of the flow $\hat u$ in the plane $x = 0$ for many different simulations. The left-most column shows simulations with $B = 1$ (effectively unstratified), the middle columns show $B = 10$ and $B=30$ (intermediate stratification) and the rightmost column shows $B = 100$ (strong stratification). The top row has $\hat S=0$, for reference, and is taken from the simulations of \citet{Garaud2020}. The remainder of the figure shows cases with $\hat S = 1$, $2$ and $4$, and two rows are shown each time: for each value of $\hat S$ the vertical shear is constant in the upper row, and sinusoidal in the lower row. Figure \ref{fig:snapshots} reveals several important trends, with increasing stratification, and with increasing shear. 

In the absence of mean vertical shear (top row, $\hat S = 0$, $J_{in} = \infty$), we see the results of \citet{Garaud2020}. For $B=1$, the fluid is effectively unstratified. The applied force drives a mean flow (visible in the snapshot) subject to horizontal shear instabilities that rapidly evolve into 3D turbulence. The latter is almost isotropic except near the forcing scale. As the stratification increases $(B=10,30)$, meanders in the streamwise flow appear (visible as a zig-zag pattern in $\hat u$). These, as discussed earlier, are driven by the horizontal shear instability and become shallower as $B$ increases. The emergent vertical shear $\partial \hat u/\partial z$ correspondingly increases and the flow adapts to be marginally unstable to a small-scale vertical shear instability, even though there is no mean vertical shear (see Section \ref{sec:resquant}). For very large stratification ($B=100$), the vertical scale of the flow meanders becomes so small that they begin to feel the effects of viscosity; the turbulence becomes intermittent. 

When $\hat S \neq 0$ but $J_{in} \gg 1$ (i.e. when $\hat S = 1$ for any $B > 1$, $\hat S = 2$ for $B = 30$ or $B = 100$, and $\hat S = 4$ for $B = 100$), we see from Figure \ref{fig:snapshots} that the snapshots are similar to those of the corresponding $\hat S = 0$ simulation, regardless of the method by which the vertical shear is forced (constant or sinusoidal). Based on the discussion of Section \ref{sec:params}, this is not surprising, as we would not expect the mean vertical shear to play a significant dynamical role, even though it is greater than the horizontal shear.  

By contrast, when $J_{in} \ll 1$ (e.g. when $\hat S = 4$ for $B \le 10$), the flow depends sensitively on the manner in which the vertical shear is forced, confirming what was found in the previous section. In the sinusoidal shear case, $\hat u$ is clearly in phase with the forcing, and is linearly unstable to the vertical shear instability, whose growth rate is $O(\hat S)$ since the stratification is weak. This is much larger than the horizontal shear instability growth rate (which is of order unity in the non-dimensionalization selected), which explains why we do not see any meanders (no zigzag patterns). 
 In the constant shear case, by contrast,  the vertical shear alone is {\it not} linearly unstable, and it is the development of flow meanders that triggers the vertical shear instability. We indeed see that $\hat u$ has strong meanders, and moreover, we now also see that these develop on roughly the same vertical scale as those of corresponding $\hat S = 0, 1$ and $2$ cases with the same value of $B$, suggesting that they arise from the same horizontal shear instability.

Finally, when $J_{in}$ is close to one, the outcome is not always as easily predictable. For instance, with $\hat S = 4$ and $B = 30$, $J_{in} = 2$ and one might have expected the flow to be dominated by horizontal shear instabilities. However, the corresponding snapshot for the sinusoidal forcing case in Figure \ref{fig:snapshots} looks much more similar to those of the simulations where $J_{in} \ll 1$, suggesting that it is probably dominated by vertical shear instabilities instead. Cases with $J_{in} \sim O(1)$ therefore require a more quantitative inspection, which is presented in the next section.

\subsection{Quantitative analysis of the results}
\label{sec:resquant}

We now take a closer look at the quantitative properties of each simulation. We first characterize the mean flow that is driven in each case and demonstrate that by constructing a Richardson number based on the actual mean vertical shear (rather than the input parameter $\hat S$), it is possible to explain the qualitative behavior of the resulting turbulent flow  more accurately. 

\subsubsection{Mean flow properties and associated regime}

In what follows, we define the mean flow to be the component of the streamwise flow $\hat u$ that projects on the particular Fourier mode whose spatial structure is the same as that of the forcing. In the case of the constant shear profile, the mean flow is defined as
\begin{equation}
    \bar U_C(y,z) = \hat U_m \sin(y) + \hat Sz, \mbox{ where } \hat U_m = \frac{\langle \hat u(x,y,z,t) \sin(y) \rangle_t}{\langle \sin^2(y) \rangle} = 2 \langle \hat u(x,y,z,t) \sin(y) \rangle_t ,\label{eq:Um1}
\end{equation}
(recalling that $\hat u$ in this case is defined as the perturbation away from the constant shear flow) so the mean vertical shear is simply $\hat S_m = \hat S$. In this expression, $\langle \cdot \rangle_t$ denotes a volume and time average over the statistically stationary phase.

 In the sinusoidal shear case, the mean flow is defined as 
\begin{equation}
    \bar U_S(y,z) = \hat U_m \sin(y+\hat Sz), \mbox{ where } \hat U_m = \frac{\langle \hat u(x,y,z,t) \sin(y+\hat Sz) \rangle_t}{\langle \sin^2(y+\hat S z) \rangle} = 2 \langle \hat u(x,y,z) \sin(y+\hat Sz) \rangle_t ,  \label{eq:Um2}
\end{equation}
so we define the quantity $\hat S_m = \hat S \hat U_m$ and also call it the mean vertical shear\footnote{Note that this terminology is used for convenience. The true vertical average of the mean vertical shear in the sinusoidal case is of course zero. By 'mean vertical shear' we imply 'the amplitude of the vertical shear of the mean flow' in this case.}.

 Using this definition, we can construct a `typical' Richardson number of the mean flow, $J_m$, based on the root-mean-square (rms) shear of the mean flow (namely $[\hat{L}_z^{-1} \int_{0}^{\hat L_z} (d\bar U/dz)^2 dz]^{1/2}$, which is $\hat S_m$ for the constant shear case, but is $\hat S_m/\sqrt{2}$ for the sinusoidal shear case): 
\begin{eqnarray}
J_m = J_C = \frac{B}{\hat S_m^2} = \frac{B}{\hat S^2} = J_{in} \mbox{ in the constant shear case} \nonumber \\
J_m = J_S = \frac{2 B}{\hat S_m^2} = \frac{2J_{in}}{\hat U_m^2}\mbox{ in the sinusoidal shear case} \label{eq:JS} .
\end{eqnarray}
Note that choosing to define $J_m$ using the rms shear of the mean flow, rather than its amplitude $\hat S_m$, eases the comparison of $J_m$ with the Richardson number of the full flow field $J_{rms}$ defined in the next section. 

\begin{figure}[!h]
  \centering
  \resizebox{0.9\hsize}{!}{\includegraphics[angle=-0]{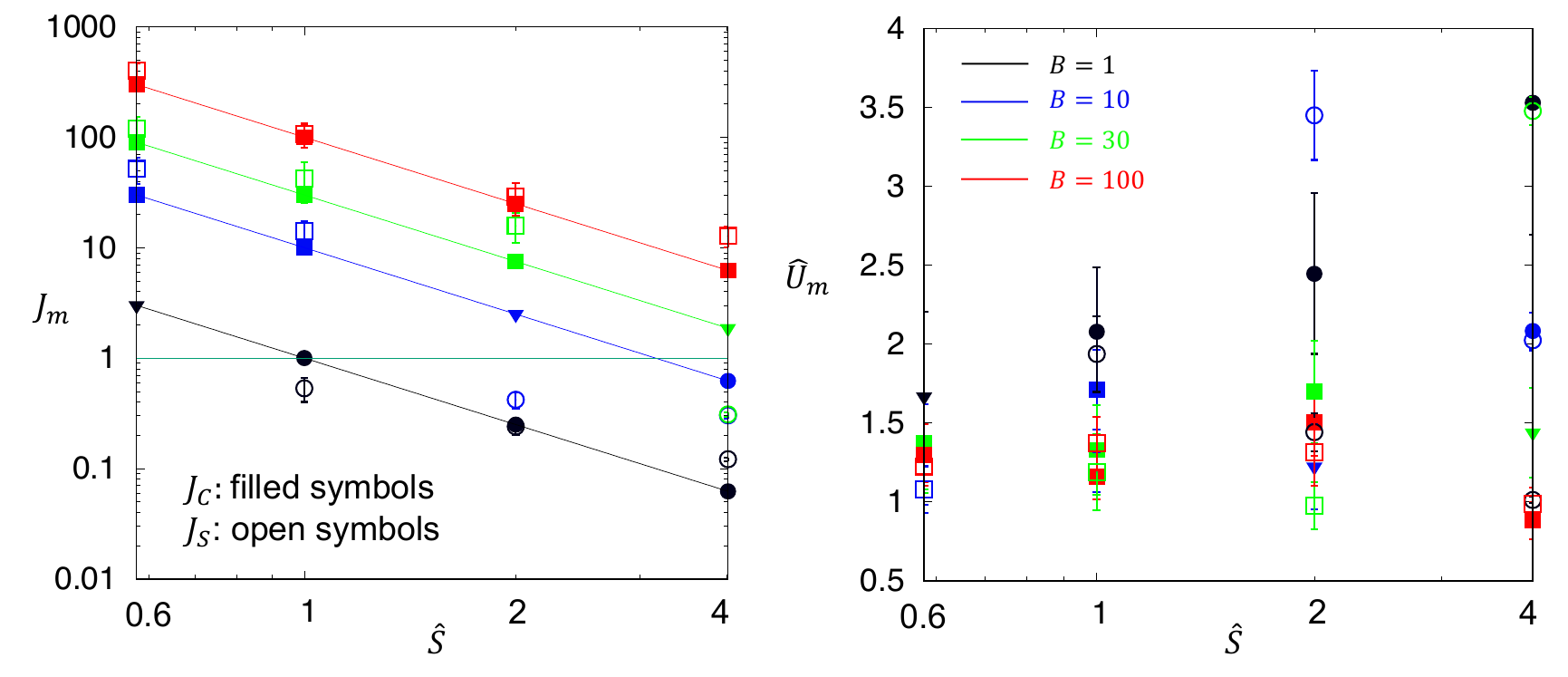}}
  \caption{Properties of the mean flow as a function of $\hat S$ and $B$. Left: typical Richardson number $J_m$ 
 of the mean flow, defined in equation (\ref{eq:JS}) ($J_C$ in the constant shear case, solid symbols, and $J_S$ in the sinusoidal shear case, open symbols). Symbol color represents $B$ as shown in the legend of the right panel. Simulations with $J_m\le 1$ are marked with circles, and are dominated by vertical shear instabilities. Those with $J_m \gg 1$ are marked with squares and are dominated by horizontal shear instabilities. Intermediate cases are marked with triangles. Right: The same for the mean flow amplitude $\hat U_m$ (see definition in equations \ref{eq:Um1} and \ref{eq:Um2}).}
\label{fig:Meanflow}
\end{figure}

The left panel of Figure \ref{fig:Meanflow} shows $J_C$ (solid symbols) and $J_S$ (open symbols) as a function of $\hat S$ for different values of $B$. These Richardson numbers measure the strength of the typical mean vertical shear relative to the imposed stratification. Using the same symbols, the right panel of Figure \ref{fig:Meanflow} shows $\hat U_m$. For the constant shear case, $J_C = J_{in} = B/\hat S^2$, which is easily predictable from the input parameters (and is shown in the straight lines / full symbols). As discussed earlier, the simulations span a wide range of Richardson numbers in the interval $[0.05,500]$. In the sinusoidal shear case, however, $J_S$ is allowed to deviate from $J_{in}$. We see that $J_S$ is close to $J_{in}$ for strongly stratified simulations, but can be substantially smaller than $J_{in}$ for more weakly stratified systems (see, in particular, $B=10$ with $\hat S = 2$, and $B=30$ with $\hat S = 4$, where $J_{in}$ is somewhat greater than 1, but $J_S < 1$). This confirms our earlier conclusion based on the snapshots only (see Figure \ref{fig:snapshots}) that these simulations are indeed dominated by vertical shear instabilities. In Figure \ref{fig:Meanflow} and in all that follows, all of the simulations that have $J_m \le 1$ are marked with solid and open circles, respectively, while those with $J_m \gg 1$ are marked with squares. We have also marked those with $J_m \in [1,4]$ with triangles, to indicate that these are borderline cases. 

To understand why $J_S$ can be much smaller than $J_{in}$ in the sinusoidal shear case, we look at the mean flow amplitude $\hat U_m$ in the right panel of
Figure \ref{fig:Meanflow}. In all strongly stratified cases (square symbols), $\hat U_m \simeq 1$. This is consistent with the dynamics being dominated by horizontal shear instabilities, where in the non-dimensionalization selected here, large-scale meanders and `pancake' eddies with horizontal lengthscale $\hat l_y \sim O(1)$ create $O(1)$ horizontal Reynolds stresses to balance the $O(1)$ forcing. For simulations dominated by the vertical shear instability (circular symbols), however, we see that $\hat U_m$ can be substantially larger. This is because the turbulent eddies have $\hat l_y \sim \hat l_z \ll 1$ in that regime \citep{GaraudKulen16,Garaud2017}. As a result, $\hat U_m$ must be correspondingly larger to ensure that the Reynolds stresses match the $O(1)$ forcing. With a large $\hat U_m$, $\hat S_m$ is also necessarily large, and $J_S$ can be substantially smaller than $J_{in}$. 

Finally, we note that this discussion is only relevant for the interpretation of the simulation results. Observations of the mean shear in real stars (or the output of stellar evolution codes that include such information) provide $\hat S_m$ directly, so there is no ambiguity on the Richardson number. In what follows, we therefore report all the remaining results in terms of $\hat S_m$ or in terms of the corresponding Richardson number $J_m$ ($J_C$ or $J_S$).  

\subsubsection{Properties of the emergent shear due to the flow meanders} 

Having identified that two regimes are expected depending on the size of the typical Richardson number of the mean flow ($J_m = J_C,J_S$), we now investigate in turn the properties of the turbulence in each case. To do so, we begin by constructing a diagnostic of the emergent shear created by the horizontal flow meanders. We first average the streamwise flow horizontally:
\begin{equation}
    \bar u(y,z,t) = \frac{1}{\hat L_x}\int_0^{\hat L_x} \hat u(x,y,z,t) dx.
\end{equation}
We then define the rms vertical shear associated with $\bar u$ as 
\begin{equation}
    \hat s_{rms} = \langle \bar s^2(y,z,t)  \rangle_t^{1/2},
\end{equation}
where $\bar s = \hat S + \partial \bar u / \partial z$ in the constant shear case, and $\bar s = \partial \bar u / \partial z$ in the sinusoidal shear case. Note that we define $\hat s_{rms}$ from the horizontally-averaged flow $\bar u$ rather than local flow $\hat u$, because a shear flow has to be coherent over some significant distance in the streamwise direction to drive vertical shear instabilities. With that definition, $\hat s_{rms} \simeq \hat S_m$ in the absence of meanders for both constant and sinusoidal shear cases, so the difference between $\hat s_{rms}$ and $\hat S_m$ can be identified as the typical meander-induced shear.

\begin{figure}[!h]
  \centering
  \resizebox{0.9\hsize}{!}{\includegraphics[angle=-0]{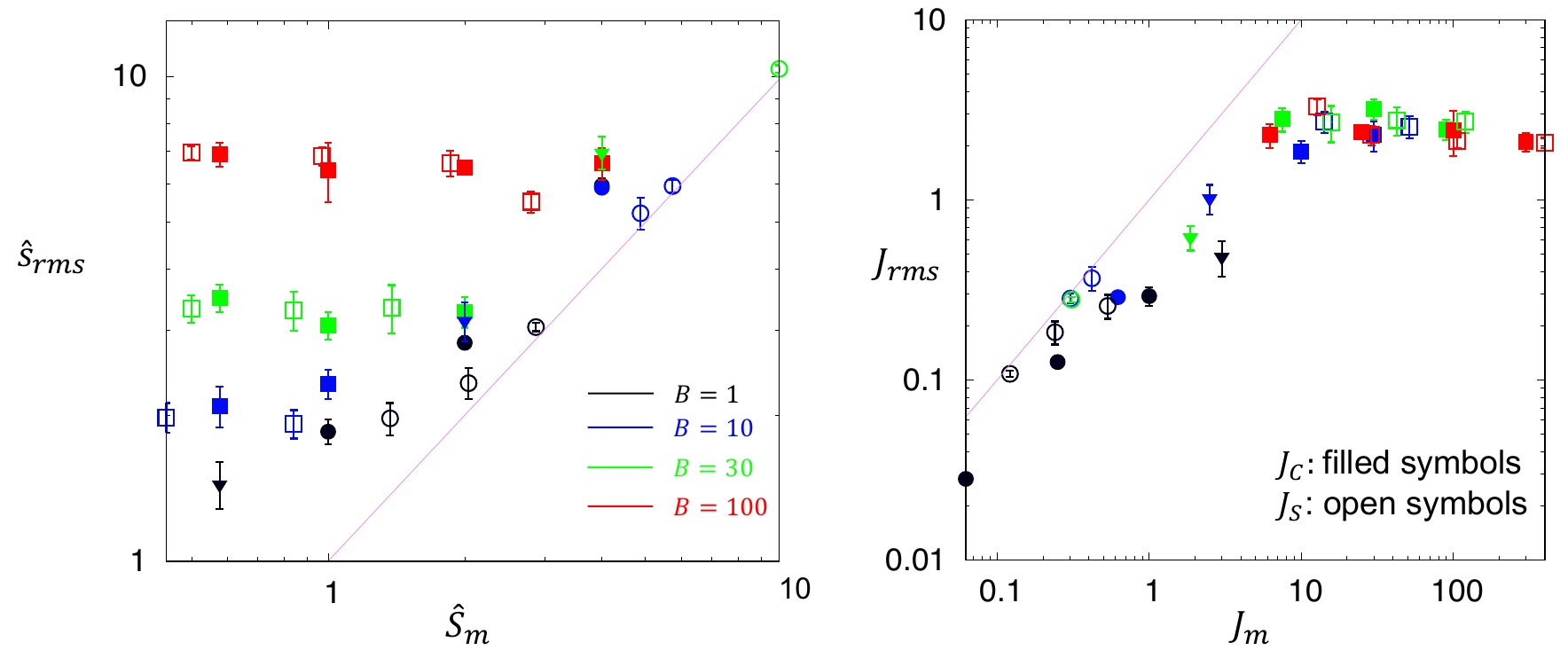}}
  \caption{Properties of the rms vertical shear emerging from the various flow instabilities. Left: $\hat s_{rms}$ as a function of the typical mean shear $\hat S_m$ ($\hat S$ in the constant shear case, solid symbols, and $\hat S \hat U_m /\sqrt{2}$ for the sinusoidal shear case, open symbols). Symbol color represents $B$. Filled symbols are for the constant shear case, and open symbols are for the sinusoidal shear case. Right: The same data presented in terms of Richardson numbers $J_{rms} = B/\hat s_{rms}^2$ vs. $J_m = B/\hat S_m^2$.}
\label{fig:srmsjrms}
\end{figure}

The values of $\hat s_{rms}$ extracted from each simulation are compared with $\hat S_m$ in Figure \ref{fig:srmsjrms} (left panel). Consistent with their definition, we always have $\hat s_{rms} \ge \hat S_m$. Very clearly, we see that simulations that are dominated by vertical shear instabilities in the sinusoidal shear case (open circles) have $\hat s_{rms} \simeq \hat S_m$, which is consistent with the visual impression originally gained from Figure \ref{fig:snapshots} that these cases do not have substantial flow meanders. In most other cases, $\hat s_{rms} \gg \hat S_m$, indicating that meanders are present and play a key role in maintaining the turbulence. In cases that are dominated by horizontal shear instabilities (i.e., $J_m\gg 1$, open and filled square symbols) $\hat s_{rms}$ seems to be independent of the mean vertical shear (and of the manner it is forced) but instead only depends on the stratification (via $B$).

 To better understand why this is the case, we now turn to the right panel of Figure \ref{fig:srmsjrms}, which shows the same information but in terms of the corresponding `rms Richardson number' 
\begin{equation}
J_{rms} = \frac{B}{\hat s_{rms}^2}
\end{equation}
against $J_m$ ($J_S$ for the sinusoidal case, $J_C$ for the constant shear case). $J_{rms}$ can be understood as a Richardson number of emergent, meander-induced, small-scale vertical shear, by comparison of $J_m$ which is a Richardson number of the large-scale vertical shear. We see very clearly that whenever $J_m \gg 1$, $J_{rms} \simeq 2$. In other words, the meanders adjust themselves precisely so that they are in a neutrally stable state with respect to the energetics of vertical shear instabilities. Indeed, it is easy to verify that if $J_{rms} \simeq 2$ then an equivalent Richardson number based on the characteristic {\it maximum} vertical shear (instead of the rms shear) would be closer to 1. When this is the case, small-scale vertical eddies extract kinetic energy from the mean shear and provide it to the background stratification in the form of potential energy in equal amounts \citep{Richardson1920} 

Dimensionally, $J_{rms} = O(1)$ implies that $s_{rms} = O(N)$. Since the horizontal velocity of the meanders is $O(U_m)$, the vertical scale of the meanders $l_z$ must be $O(U_m/N)$ to ensure that $s_{rms} = O(U_m / l_z) =O(N)$. In short, Figure \ref{fig:srmsjrms} shows that the meanders of the flow adjust themselves to be close to marginal stability for the vertical shear instability, and to do so self-consistently, must have vertical structure on scales $l_z =O(U_m / N)$. 
This characteristic scale for stratified turbulence at high P\'eclet number has long been known in the geophysical literature \citep[e.g.][]{HolfordLinden1999,BillantChomaz2001,Brethouweral2007}. It is entirely consistent (from a theoretical perspective) with the recent multiscale analysis of \citet{Chinial2022} and \citet{Shah2023}, who argued that this scaling is expected as long as $Pe \gg B^{1/2}$ and as long as viscous effects can be neglected. Our results suggest that it continues to hold as long as the Richardson number based on that mean vertical shear is substantially larger than one (e.g., $J_m \ge 5$, say). 

For weaker stratification, namely $J_m \le O(1)$, we saw in Section \ref{sec:qualtemp} that the statistically stationary state achieved depends on the manner in which the flow is forced. For the sinusoidal case (open circles), we saw that meanders are essentially absent, so $J_{rms} \simeq J_m$. In the constant-shear case (filled circles and triangles), on the other hand, the mean flow is linearly stable to the vertical shear instability, so the latter relies on the onset of meandering via horizontal shear instabilities to set in. In that case, $\hat s_{rms}$ must necessarily exceed $\hat S_m$. In practice, however, we find that $\hat s_{rms}$ is only a little larger than $\hat S_m$ or equivalently, $J_{rms}$ is a little smaller than $J_m$. 

\subsubsection{Turbulent mixing}
\label{sec:turbmix}

We now study vertical mixing by the turbulence in each simulation. As introduced in Section \ref{sec:intro}, an important quantity of interest for stellar evolution calculations is the so-called vertical mixing coefficient $D_{mix}$, which is the turbulent diffusivity of momentum or a trace chemical element. As long as the flow is not in a viscously influenced regime where the turbulence is intermittently suppressed, this coefficient can be estimated (non-dimensionally) from the product of a characteristic vertical velocity of the turbulence ($\hat{w}_{rms}$) and a characteristic vertical eddy height ($\hat{l}_z$):
\begin{equation}
\hat D_{mix} \simeq c \hat{w}_{rms}\hat{l}_z  \rightarrow D_{mix} \simeq c \hat{w}_{rms}\hat{l}_z U_h L_h
\end{equation}
where $c$ is some constant of order unity. 
The quantity $\hat{w}_{rms}$ is easily obtained from the simulations by taking the time average of $\hat{w}_{rms}(t)$ once the system has reached a statistically stationary state. For the one shown in Figure \ref{fig:B10S1}, for instance, that average is taken between $t = 30$ and $t = 95$, and is equal to $\hat{w}_{rms} = 0.59 \pm 0.03$, where the $\pm 0.03$ captures the standard deviation of $\hat{w}_{rms}(t)$ around the mean value $0.59$. The values of $\hat{w}_{rms}$ thus computed, for all available simulations, are presented in Table 1. 

To calculate the eddy height $\hat{l}_z$, following \citet{Garaud2020}, we compute the vertical auto-correlation of the vertical velocity field, defined as:
\begin{equation}
A_w(\hat l,t) = \langle \hat w(x,y,z,t) \hat w(x,y,z+\hat l,t) \rangle \, .
\label{eq:Aw}
\end{equation}
The vertical lengthscale of the eddies at a given time $t_n$ is then defined as the width at half maximum of $A_w(\hat l,t_n)$. In other words,  $\hat{l}_z(t_n)$ is the solution of 
\begin{equation}
A_w(\hat{l}_z(t_n),t_n) = \frac{1}{2} A_w(0,t_n)\, .   
\end{equation}
Finally, $\hat{l}_z$ is defined as the average of all lengthscales thus computed for a given simulation once the system is in the statistically stationary state. The results are presented in Table 1. Note that the full velocity field is only saved every 10,000 timesteps at this value of $Re$, which, owing to the Courant condition, approximately corresponds to intervals $\Delta t_n$ of 0.5 to 3 time units depending on the simulation. In general $\hat{l}_z$ is then computed from an average of 20--60 snapshots. Because this number is relatively small, the statistics of $\hat{l}_z$ are not as robust as those of $\hat w_{rms}$.  

\begin{figure}[!h]
  \centering
   \resizebox{0.9\hsize}{!}{\includegraphics[angle=0]{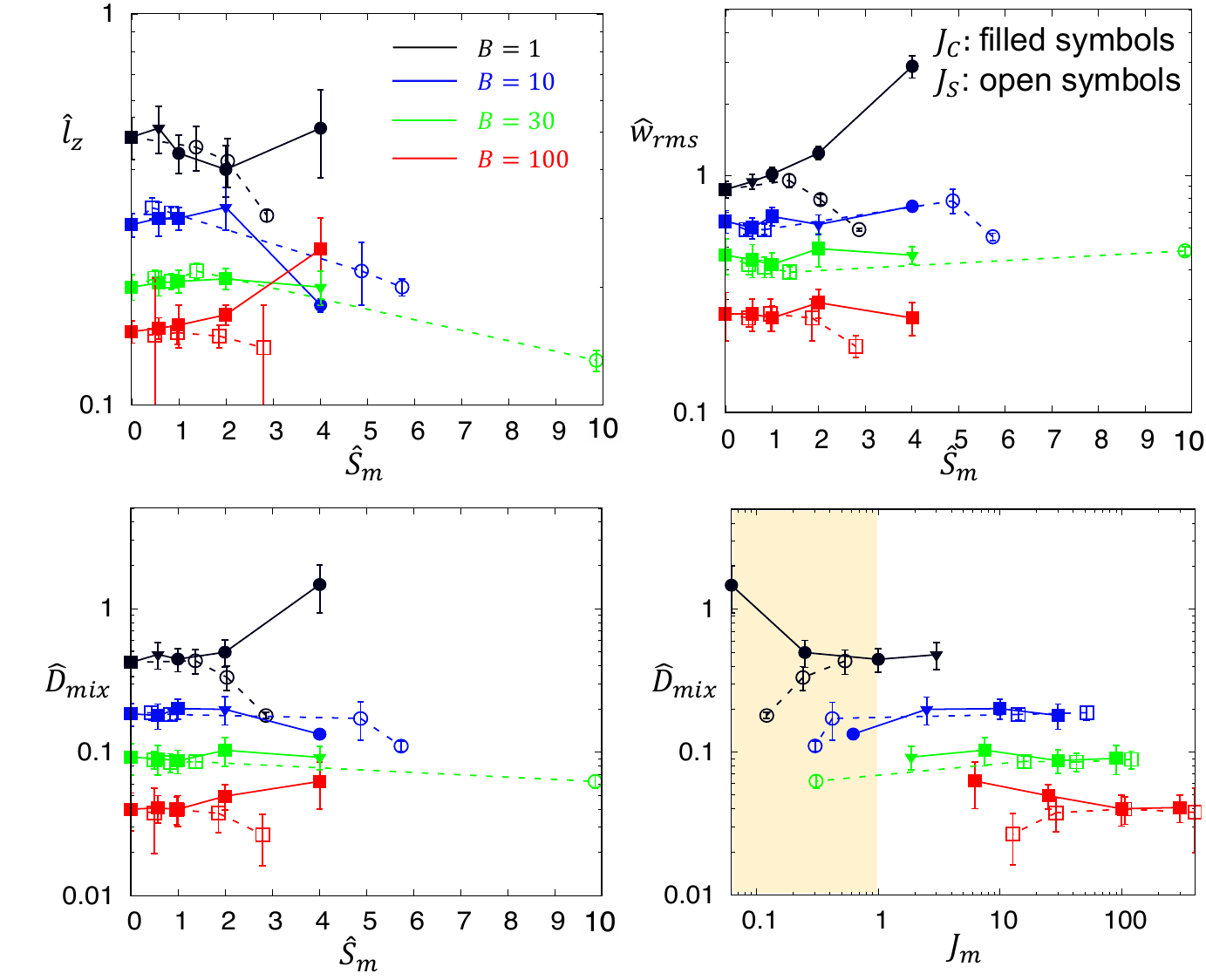}}
  \caption{Vertical eddy scale $\hat l_z$ 
  (top left), rms vertical velocity $\hat w_{rms}$ (top right), and estimated mixing coefficient $\hat D_{\rm mix} = \hat l_z \hat w_{\rm rms}$ (bottom left) as a function of the typical mean vertical shearing rate $\hat S_m$ and the stratification $B$ (indicated by the line colors, see legend). Filled symbols connected by solid lines correspond to the constant shear case, and open symbols connected by dashed lines correspond to the sinusoidal shear case. The bottom right figure shows $\hat D_{\rm mix}$ as a function of the typical mean Richardson number $J_m =(J_C,J_S)$}
\label{fig:results}
\end{figure}

Fig. \ref{fig:results} shows $\hat l_z$ (top left) and $\hat w_{rms}$ (top right) as a function of the typical vertical shearing rate of the mean flow $\hat S_m$ for various values of the stratification parameter $B$. Consistent with the qualitative picture revealed by Figure \ref{fig:snapshots} and originally presented in \citet{Garaud2020} for the $\hat S_m = 0$ dataset, we see that both the vertical eddy scale and the vertical velocity decrease with increasing stratification (increasing $B$). As $\hat S_m$ increases 
at fixed values of $B = 1, 10$ and $30$ (but not $B=100$, see below), we see that the mean shear, and the manner in which it is forced, have little influence on $\hat l_z$ and $\hat w_{rms}$ as long as $J_m \ge 1$ (open and closed square and triangular symbols), but begin to affect them when $J_m < 1$. Overall, this is not surprising based on the discussions of the previous sections. By contrast, the effects of 
 a strong ($J_m < 1$) constant or sinusoidal shear on $\hat l_z$ and $\hat w_{rms}$ are complex (sometimes increasing them, sometimes decreasing them). Since this limit is fairly rare in stellar interiors, we do not investigate it further in this paper.  

In the strongly stratified case ($B=100$), we were initially surprised to find that the mean shear does affect the turbulence even though $J_m \gg 1$ -- in this case, it does so as soon as $\hat S_m > 1$. 
However, this behavior can be understood by noting that $B=100$ is in a viscously influenced regime where the turbulence is spatially patchy at $\hat S= 0$ \citep{Garaud2020}. When that is the case, the fraction of the domain covered by turbulent patches depends quite sensitively on the existence of nonlinear mechanisms that help the turbulence self-sustain and percolate through the flow \citep{Deusebioal2015,Avilaal2023}. We hypothesize that a strong vertical shear might stretch these turbulent patches and disrupt the self-sustaining process. The effect is clearly different in the constant and sinusoidal shear cases, sometimes increasing or decreasing $\hat l_z$ or $\hat w_{rms}$. We defer the analysis of the intermittent regime to future work, as we believe that it is not particularly relevant to stellar interiors. Indeed \citet{Shah2023} demonstrated that stratification is strong enough for viscosity to influence the turbulence only when $B > Re$ (unless thermal diffusion is important, see the original paper for detail), or equivalently in dimensional terms when 
\begin{equation}
\frac{U_h}{L_h} \ll \left(\frac{N^2 \nu}{L_h^2}\right)^{1/3}.
\label{eq:turbulentcrit}
\end{equation}
 In stars, this corresponds to regions of very small horizontal shear, in which the corresponding vertical transport would likely be negligible anyway. 

The lower two panels of Figure \ref{fig:results} show $\hat D_{mix} = \hat w_{rms} \hat l_z$ as functions of $\hat S_m$ (left) and $J_m = J_C, J_S$ (right). The values of $\hat D_{mix}$ for $B = 100$ are shown for reference, but should be ignored as the formula used is not reliable in that case because of the patchiness of the turbulence. Following our findings for the behavior of $\hat l_z$ and $\hat w_{rms}$, we see that $\hat D_{mix}$ decreases steadily with increasing stratification and is roughly independent of the shearing rate and the manner in which the vertical shear is forced as long as $J_m \ge 1$ (squares and triangles). 

Based on the results presented in this section we can therefore make the following statement with reasonable confidence: as long as thermal diffusion and viscosity are negligible (so the turbulence is both fully developed and adiabatic), a mean vertical shear has no effect on mixing driven by horizontal shear instabilities when its mean Richardson number is  greater than one. While this is not particularly surprising in hindsight, it is worth emphasizing that this statement likely applies to the vast majority of stellar shear layers, {\it even when the vertical shearing rate is much greater than the horizontal one}, which is perhaps a little less intuitive. A good example of this is the solar tachocline, where the vertical shear is over 20 times stronger than the horizontal shear \citep{Charbonneaual1999}. The fact that $J = O(10^3 - 10^4) \gg 1$ in the tachocline \citep{Garaud2020} shows that unless horizontal shear instabilities are somehow suppressed by processes not included here (see the discussion below), the turbulence they generate dominates vertical mixing, while the mean vertical shear can essentially be ignored in the computation of $D_{mix}$!  

\section{Discussion and Conclusions} \label{sec:conclusions}

In this paper, we have used DNS to investigate the combined effects of vertical and horizontal shear on mixing by shear instabilities in the stably stratified regions of stars. We ignored rotation and magnetic fields for now. The horizontal shear was forced to be sinusoidal in the spanwise direction (here, $y$), and we compared two types of vertical shear profiles (constant vs. sinusoidal). The numerical simulation parameters were selected to be in a regime where thermal diffusion is negligible on the large horizontal and vertical scales, while having a low Prandtl number. This regime (rather than the numerical parameters themselves) is consistent with what is expected in a typical stellar interior \citep{Shah2023}.

Within this set of assumptions, we found that two dynamical regimes naturally emerge, depending on the Richardson number $J_m = N^2 /S_m^2$, where $S_m$ is a typical value of the large-scale vertical shear. When $J_m$ is of order unity or smaller, vertical shear instabilities dominate the system dynamics. The form they take (and the amount of mixing they cause) depends sensitively on the geometry of the vertical shear and more specifically on whether it can trigger a linear instability or not. In addition, even though we have only considered regions with a constant stratification in this paper, there is ample evidence from the oceanographic literature \citep[see the review by][]{Caulfield2021} showing that the outcome will depend sensitively on the shape of the density profile and must be studied on a case-by-case basis. Regions of stars with $J_m < O(1)$ are rare, however, so this regime is not particularly relevant to stellar evolution in general. 

In the far more likely instances where $J_m \gg 1$, horizontal shear instabilities control the system dynamics, and the mean vertical shear has little influence on vertical mixing (unless the stratification is so strong that the turbulence itself is viscously suppressed, see Section \ref{sec:turbmix}). This is because the large-scale horizontal perturbations excited by the horizontal shear instability (meanders, or pancake vortices) decouple in the vertical direction on small vertical scales and drive a much stronger vertical shear. More specifically, we showed in Section \ref{sec:resquant} that this emergent meander-induced shear $s_{rms}$ satisfies the marginal stability condition $J_{rms} = N^2 / s^2_{rms} = O(1)$, so $s_{rms} = O(N)$, which by definition of this regime is much larger than the mean shear $S_m$. As a result, vertical mixing is entirely driven by the emergent rather than mean vertical shear. When that is the case, models for mixing by horizontal shear instabilities alone can be safely applied -- at least once the controversy as to which model should be used is resolved, see Section \ref{sec:introhoriz} for detail.  

There are, of course, several caveats to this statement. First, note that we have ignored in this work the effects of thermal diffusion by purposefully selecting parameters in which the turbulence generated is known to be relatively adiabatic. This, we believe \citep[see][]{GaraudKulen16,Garaud2020,Shah2023}, is the more common situation in stars, but that belief is somewhat controversial \citep[see, e.g.][]{Skoutnev2023}. Presumably, the criterion will be similar in the thermally diffusive case -- i.e the mean shear can be ignored when it is much smaller than the meander-induced shear -- but this hypothesis will need to be verified. More importantly, we have also ignored in this paper several effects that could suppress the horizontal shear instabilities themselves. Indeed, sufficiently strong rotation, or strong streamwise magnetic fields (both of which are present in the solar tachocline, for example) can stabilize purely two-dimensional horizontal shear instabilities \citep[see, e.g., the discussion in][]{Garaud2021}. When that is the case, the vertical shear may be needed to provide an alternative route to turbulence and must not be neglected.

Finally, we acknowledge that there are, of course, many other forms of hydrodynamic or magnetohydrodynamic (MHD) instabilities in stars, such as double-diffusive thermocompositional instabilities \citep[see the reviews by][]{Garaud2018AnRFM,GaraudDDreview2020}, centrifugal instabilities \citep{solberg_mouvement_1936,hoiland_interpretation_1939}, their double-diffusive counterpart \citep[namely the GSF instability][]{goldreich_differential_1967,barker_angular_2019,barker_angular_2020}, baroclinic instabilities \citep{SpruitKnobloch1984,GilmanDikpati2014}, magnetic buoyancy instabilities \citep{Acheson1979,SpiegelWeiss1982,VasilBrummell2008}, MHD shear instabilities \citep{DikpatiGilman1999,Spruit1999}, the magnetorotational instability \citep{BalbusHawley1991,MenouLemer2006} and the Tayler-Spruit instability \citep{Tayler1973,Spruit2002,Jial2023,Petitdemange2023}. These (and others) are known to be relevant at various evolutionary stages, and will take over if their growth rate significantly exceeds that of the horizontal shear instabilities. As such, this work is only relevant when these other sources of mixing are negligible. 

Nevertheless, by carefully characterizing the conditions under which each instability can grow, and how much mixing they cause, we can gradually make progress in advancing our understanding of their role in stellar evolution and parametrize their effect in 1D \citep[e.g. MESA,][]{Paxton2013} and now 2D \citep[cf. ESTER,][]{Mombarg2023} stellar evolution codes. This paper has taken one more step in this direction. 

%\begin{acknowledgments}
%This work was initiated as part of the Kavli Summer Program in Astrophysics hosted virtually at the Max Planck Institute for Solar System Physics in G\"ottingen in 2021, funded from a generous grant by the Kavli Foundation. P.G. acknowledges funding by NSF AST-1814327. J.M.B. acknowledges funding by NSF OCE-1756491. We thank the referee for their help in improving the clarity of the paper. 
%\end{acknowledgments}

\bibliography{dns_shear}{}
\bibliographystyle{aasjournal}

\end{document}